\documentclass[a4paper,fleqn]{cas-dc}

\usepackage[authoryear,longnamesfirst]{natbib}
\usepackage{subcaption}
\usepackage{threeparttable}

\def\tsc#1{\csdef{#1}{\textsc{\lowercase{#1}}\xspace}}
\tsc{WGM}
\tsc{QE}
\tsc{EP}
\tsc{PMS}
\tsc{BEC}
\tsc{DE}

\begin{document}
\let\WriteBookmarks\relax
\def\floatpagepagefraction{1}
\def\textpagefraction{.001}

\shorttitle{Improving Student-AI Interaction Through Pedagogical Prompting}

\shortauthors{R. Xiao, et al.}

\title [mode = title]{Improving Student-AI Interaction Through Pedagogical Prompting: An Example in Computer Science Education}

\author[1]{Ruiwei Xiao}[orcid=0000-0002-6461-7611]\cormark[1]
\author[2]{Xinying Hou}[orcid=0000-0002-1182-5839] \cormark[2]
\author[3]{Runlong Ye}[orcid=0000-0003-1064-2333] \cormark[2]
\author[3]{Majeed Kazemitabaar}[orcid=0000-0001-6118-7938] \cormark[2]
\author[4]{Nicholas Diana}[orcid=0000-0002-8187-3692]
\author[5]{Michael Liut}[orcid=0000-0003-2965-5302]  
\author[1]{John Stamper}[orcid=0000-0002-2291-1468]

\affiliation[1]{organization={Carnegie Mellon University},
    city={Pittsburgh},
    state={PA},
    country={United States}
}

\affiliation[2]{organization={University of Michigan},
    city={Ann Arbor},
    state={MI},
    country={United States}
}

\affiliation[3]{organization={University of Toronto},
    city={Toronto},
    state={ON},
    country={Canada}
}

\affiliation[4]{organization={Colgate University},
    city={Hamilton},
    state={NY},
    country={United States}
}

\affiliation[5]{organization={University of Toronto Mississauga},
    city={Mississauga},
    state={ON},
    country={Canada}
}

\cortext[cor1]{Corresponding author. Email: ruiweix@cs.cmu.edu.}
\cortext[cor2]{These authors contributed equally.}

\begin{abstract}
With the proliferation of large language model (LLM) applications since 2022, their use in education has sparked both excitement and concern. Recent studies consistently highlight students' (mis)use of LLMs can hinder learning outcomes. This work aims to teach students how to effectively prompt LLMs to improve their learning. We first proposed pedagogical prompting, a theoretically-grounded new concept to elicit learning-oriented responses from LLMs. To move from concept design to a proof-of-concept learning intervention in a real educational setting, we selected early undergraduate CS education (CS1/CS2) as the example context. We began with a formative survey study with instructors teaching early-stage undergraduate-level CS courses (N=36) to inform the instructional design based on classroom needs. Based on their insights, we designed and developed a learning intervention as an interactive system with scenario-based instruction to train pedagogical prompting skills. Finally, we evaluated its instructional effectiveness through a user study with undergraduate CS novices (N=22) using pre/post-tests. Through mixed methods analyses, our results indicate significant improvements in learners’ LLM-based pedagogical help-seeking skills along with positive attitudes toward the system and increased willingness to use pedagogical prompts in their future learning. The contributions include (1) a theoretical framework of pedagogical prompting; (2) empirical insights into current instructor attitudes toward pedagogical prompting; and (3) a learning intervention design with an interactive learning tool and scenario-based instruction leading to promising results on teaching LLM-based help-seeking. Our approach is scalable for broader implementation in classrooms and has the potential to be integrated into tools like ChatGPT as an on-boarding experience to encourage learning-oriented use of generative AI.
\end{abstract}



\maketitle

\section{Introduction}
Students may face difficulties in their learning and need to seek instructional support \citep{bornschlegl2020variables}. Such help-seeking is especially common in domains like computer science education, which often present significant challenges to novices due to their steep learning curve \citep{zahn2024investigating,cohen2024factors}. While learners have traditionally relied on office hours, peers, or online resources for help-seeking \citep{wirtz2018resource}, the widespread availability of large language models (LLMs) has shifted many students toward using these AI tools instead. However, this increased availability can exacerbate existing help-seeking issues \citep{prather2024widening} and even introduce new challenges, such as over-reliance \citep{kazemitabaar2023novices}, superficial engagement, or failing to identify incorrect responses. The unproductive LLM-based help-seeking can be correlated with declining learning outcomes compared to students before LLM became prevalent \citep{Osorio2025} or students in controlled groups without access to LLMs \citep{Jost2024}. 

\begin{sloppypar}To promote more learning-oriented use of LLMs, researchers have proposed tools that embed pedagogical guardrails \citep{liffiton2023codehelp, kazemitabaar2024codeaid, xiao2024exploring,hou2024codetailor}. These tools aim to provide programming support without revealing direct solutions \citep{pirzado2024navigating}. While such designs encourage deeper engagement, they often constrain the full capabilities of LLMs and fall short of aligning with learners' expectations or real-world needs. As a result, learners tend to underutilize these systems, favoring more versatile tools like ChatGPT, even if these are used in ways that hinder long-term learning \citep{kazemitabaar2024codeaid, xiao2024preliminary}. In contrast, while general-purpose LLMs are capable of pedagogical support, students often do not use them pedagogically for learning. Instead, many students take advantage of this ``flexibility'' to obtain direct assignment answers, leading to a false sense of competence and delayed recognition of learning gaps \citep{prather2024widening}. Given that LLMs are becoming an increasingly important part of students’ educational and professional experiences, it is critical to proactively teach students how to use general-purpose LLM tools, especially AI chatbots, to facilitate their learning.
\end{sloppypar}

Motivated by a commitment to promote student agency \citep{mercer2011understanding} and enable 24/7 accessible learning across AI tools, we focus on teaching students prompting-for-learning skills. That is, the ability to craft prompts that guide LLMs to function as effective personal tutors and deliver appropriate instructional strategies. In this way, learners are be able to engage with these tutoring LLMs throughout the learning process until mastery. To achieve this goal, in this study, we proposed the \textbf{\textit{concept of pedagogical prompt}}, a class of prompts strategically designed to guide AI responses in alignment with instructional approaches (e.g., worked example, prompted self-explanation) suited to a learner's specific struggle. To move from conceptual design to a proof-of-concept in real educational settings, we chose early undergraduate computer science (CS) education as our initial focused context. We first conducted a \textbf{\textit{formative study}} with instructors of early undergraduate computer science courses (CS1 and CS2) to identify their preferred instructional delivery methods for teaching pedagogical prompting skills. Based on insights from 36 instructors, the study confirmed the need to teach pedagogical prompts in early undergraduate CS classrooms and informed the design of an interactive system to deliver instruction. We then developed this interactive system based on instructors' insights and employed scenario-based instruction within the context of early CS education. To evaluate the impact of this intervention on teaching pedagogical prompting, we conducted a \textit{\textbf{user study}} with 22 students who recently completed their first undergraduate CS course.

Findings from our study indicate significant learning gains on writing pedagogical prompts from pre-test to post-test, along with an increased perceived ability to effectively use LLM tools for learning. Additionally, participants rated the learning experience positively and expressed a strong willingness to continue using pedagogical prompts in their future learning. Building on this initial success, our implementation has the potential to scale for broader adoption in classrooms and be integrated into commercialized AI platforms like ChatGPT as an on-boarding experience. Such an integration would foster learning-oriented LLM interactions across different learning contexts and would benefit students both within and beyond computing education.
\section{Related Works}
This section includes work about (1) students' existing limitations in different help-seeking stages, illustrated through the context of CS education (2.1); (2) current solutions to improve help-seeking effectiveness with LLMs in CS education and their associated challenges (2.2); and (3) the main theory to link learning context with appropriate AI tutoring activity in pedagogical prompts (2.3).

\subsection{Current limitations of students' help-seeking with LLM in CS education}
\begin{figure*}
    \centering
    \includegraphics[width=1\linewidth]{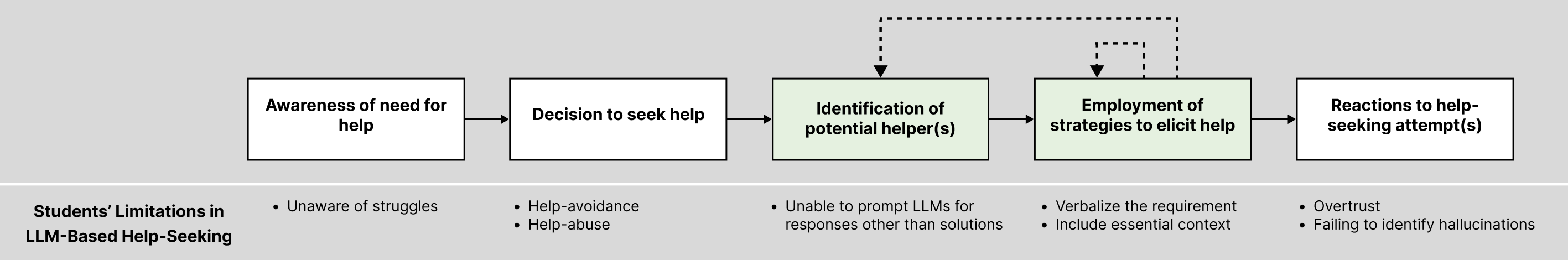}
    \caption{Original model of help-seeking process (top-half) and student limitations in (LLM-Based) help-seeking (bottom-half). Blocks in Green (Step 3 and 4): Our intervention directly focuses on.}
    \label{fig: help-seeking}
\end{figure*}

Help-seeking has been widely studied by educational researchers. \citet{nelson1981help} defined help-seeking as \textit{"the attempts to obtain assistance or intervention from another person when they cannot gratify their needs or attain a desired goal through their own efforts"} and proposed a 5-stage model of the help-seeking process (Figure \ref{fig: help-seeking} upper-half). As a meta-cognitive skill for learning, help-seeking been strongly linked to students’ performance in CS \citep{marwan_unproductive_2020, bergin2005examining, biro2014deep}. Despite its importance, difficulties in seeking help effectively have been a persistent challenge in CS education \citep{marwan_unproductive_2020} and across domains \citep{aleven_help_2003}. Understanding existing challenges is essential for designing instructional interventions that foster productive help-seeking behaviors in novice programmers.

Before the existence of LLM-based help-seeking, in CS education, researchers have identified factors that influence help-seeking in computer-based learning environments and with humans (tutor, peers, etc.) \citep{price2017factors}, patterns of help-seeking (e.g., tendency to use high-accessibility but low-utility helpers) \citep{doebling2021patterns} and problematic help-seeking behaviors including help-abuse and help-avoidance \citep{marwan_unproductive_2020}. While LLMs enhance the flexibility of help-seeking and broaden the range of potential helpers, it also introduced challenges for help-seeking. Repeated patterns demonstrate that ineffective help-seeking with LLMs can hinder learning gains \citep{padiyath2024insights}. For example, \citet{Osorio2025} found that those who use commercially available LLMs (e.g., ChatGPT) for assignment problems performed better on those problems, but performed worse on the overall assessments and even in the course. Similarly, \citet{Jost2024} observed a significant negative correlation between increased reliance on LLMs for critical thinking-intensive tasks, such as code generation and debugging, and lower final grades in programming education. Both \citet{Osorio2025} and \citet{Jost2024} found that LLM usage correlates with decreased student performance, indicating that excessive dependence on LLM assistance could impair the development of essential independent problem solving skills. However, \citet{Kumar2024} highlighted the potential for LLMs to positively influence learner performance and engagement when thoughtfully integrated into educational contexts. For example, providing structured guidance, such as explicit instructions or prompting students to first attempt problems independently before refining solutions with LLM assistance, can reduce unproductive queries and enhance student engagement, ultimately fostering greater trust and confidence in both the technology and their own problem solving abilities. More specifically, learners exhibit less-regulated behaviors throughout the following stages in the help-seeking model during LLM-based help seeking, which potentially contribute to poor learning outcomes: 

\textit{\textbf{Stage 1}: Awareness of need for help}. This stage is not specific to a particular helper, thus the fundamental challenges of traditional help-seeking remain. Furthermore, LLM-based chatbots (e.g., ChatGPT) require users to initiate interactions rather than having system-initiative or mixed-initiative \citep{horvitz1999principles} mechanism (e.g. \citep{marwan2020adaptive}). Thus, recognizing the need for help and clearly defining one’s struggle are vital in LLM-based help-seeking.

\textit{\textbf{Stage 2}: Decision to seek help}. As an additional stage not specific to helpers, LLM-related literature reported a continued, undesired pattern for both help-abuse \citep{kazemitabaar2023novices, padiyath2024insights} and help-avoidance \citep{padiyath2024insights} with various reasons. For example, \citet{kazemitabaar2023novices} observed CS1 students' over-reliance (e.g., too frequent help request) on LLMs during problem-solving. Regarding help-avoidance, \citet{padiyath2024insights} found a negative correlation between students’ self-reported use of LLMs in class and their concern that over-reliance on LLMs harms their programming skills. Either the over-trust or over-concerns can affect learners' decision to seek help from LLMs. 

\textit{\textbf{Stage 3}: Identification of potential helper(s)}. Moreover, many students fail to identify LLMs as a tutor but perceive LLMs as (direct, correct) solution provider. In one-third of the cases studied by \citet{amoozadeh2024student}, students submitted the full task to ChatGPT for the direct solution without any efforts on their own. This finding suggests that few students explicitly treat LLMs as instructional tutors and use LLMs in a pedagogical way accordingly.

\textit{\textbf{Stage 4}: Employment of strategies to elicit help}. Even when learners are aware of treating LLMs as tutors, they often fail to articulate their needs clearly and thus do not receive the intended responses. \citet{nguyen2024beginning} identified barriers such as difficulties in expressing problem and using appropriate coding terminology. Using data from a large-scale deployment of an LLM tutor in intro-level CS classrooms, \citet{xiao2024preliminary} found that novice CS students were incompetent in writing good prompts with a clear request and essential context \citep{xiao2024preliminary}, indicating students' struggle in \textit{\textbf{Stage 4}} to construct the help request as perceived by the help provider \citep{langer1972semantics,nelson1981help}. 

\textit{\textbf{Stage 5}: Reactions to help-seeking attempt(s)}. When receiving responses from an LLM, novices often lack awareness or face challenges in evaluating the accuracy and reliability of the information provided. They often lack verification strategies \citep{amoozadeh2024student}, tend to over-trust the AI’s outputs \citep{kazemitabaar2023novices}, and struggle to identify hallucinations, such as misleading or incorrect responses generated by the model \citep{xiao2024preliminary}. These can lead to the uncritical adoption of incorrect solutions, reinforcing misconceptions and hindering learning. 

This work focuses on two uniquely important stages that set up the ground for the student-AI interaction. The first, \textit{\textbf{Stage 3}: identification of potential helpers} aims to increase learners’ awareness of LLMs as viable pedagogical support tools. Secondly, \textit{\textbf{Stage 4}: employment of strategies to elicit help} focuses on strengthening learners’ ability to verbalize their needs effectively when constructing prompts. While we acknowledge the importance of other stages, \textbf{\textit{Stage 1}} and \textbf{\textit{Stage 2}} are more general cognitive steps that are not unique to AI-mediated environments and often precede tool-specific interactions; \textbf{\textit{Stage 5}} is important and influenced by AI, however, it often depends on the quality of interactions in \textbf{\textit{Stage 3}} and \textbf{\textit{Stage 4}}, thus becomes less open to refine unless earlier student-AI interaction is improved first. Therefore, this work focuses on \textbf{\textit{Stages 3}} and \textbf{\textit{Stage 4}}, with \textit{\textbf{Stage 5}} remaining a potential direction for future research based on the findings of this work.

\subsection{Existing pedagogical LLM tools in CS education}
Over the past three years, a large number of LLM-based tutoring systems have been developed to assist novices for their programming learning \citep{stamper2024enhancing, liffiton2023codehelp, liu2024teaching, kazemitabaar2024codeaid, kumar2024supporting, schmucker2024ruffle, xiao2024exploring, hou2024codetailor, hou2025personalized}. With extensive efforts, many of these systems iterated from a naive integration of chatbot into CS learning context, to carefully integrated learning science principles into prompt design for generating more pedagogical help. For example, there are guardrail tools that provide predefined hints or a certain type of guidance (e.g., learning-by-teaching \citep{schmucker2024ruffle}, self-reflection \citep{kumar2024supporting}, worked examples \citep{xiao2024exploring}, practice problems \citep{hou2024codetailor}). Although incorporating learning science guidelines into predefined prompts helps prevent learners from directly copying answers, it also limits the access to adaptive support for different problems and has a weaker impact on fostering flexible help-seeking behaviors and enhancing students' ability to seek help effectively. A limited number of studies focus on prompt engineering skills for programming, with most concentrating on crafting prompts to generate correct code \citep{denny2023promptly}. As a result, despite previous efforts in the pedagogical LLM-based tools, there remains a gap between supporting learners to develop more structured and pedagogically sound help-seeking behaviors when using LLMs.

\subsection{Link learning context to LLM-based tutoring activity: KLI Framework as the guideline}
With the lessons learned in the past years, the community generally agrees that when asking for LLMs for responses, directly providing solutions is an ineffective help-seeking strategy. However, what constitutes an effective prompt to learning remains underexplored \citep{martin2021help}. In the context of CS education, while specifying that the LLM should "behave as a programming tutor" can signal an instructional intent, it does not clarify the expected LLM response format, making the content of responses and level of pedagogy difficult to control. To address this ambiguity, it is essential not only to define the role of the LLM, but also to clearly specify the tutoring protocol, that is, the expected structure and style of the instructional interaction. Prior research, such as the Knowledge-Learning-Instruction (KLI) framework \citep{koedinger2012knowledge} in learning sciences suggests that teachers' choice of instructional method should depend on the type of knowledge the student needs to acquire. The KLI framework categorizes different types of knowledge and maps them to corresponding learning events and suitable instructional methods. In this work, we use the KLI framework to guide the selection of tutoring protocols within pedagogical prompts, ensuring alignment between the knowledge type being acquired and the most effective tutoring strategies.

The KLI framework introduces different types of knowledge components (KCs, acquired units of skill that can be inferred through
performance on a set of problems) and categorizes three broad classes of learning events (LEs): (a) memory and fluency processes, (b) induction and refinement processes, and (c) understanding and sense-making processes. It then establishes how these learning events are mapped to evidence-based instructional methods by specifying the optimal instructional strategies for each type of learning process. For example, \textit{memory and fluency processes} benefit from techniques such as retrieval practice and spaced repetition, \textit{induction and refinement processes} are supported by guided discovery and worked examples, and \textit{understanding and sense-making processes} are best facilitated through self-explanation, inquiry-based learning, and interactive discussions. For example, code writing (e.g., using for loop to iterate over a list of numbers) can be seen as a rule type of KC, as it is a one-to-many mapping, therefore induction and refinement is the most effective learning process, and worked example \citep{sweller1985use} is one of the instructional principle that can effectively teach such knowledge. Similar analysis can be applied to debugging, which falls into the principle type of knowledge hence understand and sense-making learning event and is the most effective and can be taught with prompted self-explanation \citep{chi1994eliciting, hausmann2007explaining}.

One of the knowledge types included in this work is code writing (e.g., completing the code of using a for loop to iterate over a list of numbers). According the definition in the KLI framework \citep{koedinger2012knowledge}, this can be categorized as a variable-variable knowledge component (KC) since it requires applying a general looping structure to different contexts. As a result, induction and refinement is the most effective learning process for acquiring this knowledge. One effective approach is the worked example method \citep{sweller1985use}, which helps learners generalize patterns and refine their understanding through guided practice. Take debugging as another example, debugging aligns with the principle type of KC as it involves diagnosing and reasoning about underlying program behavior. Because debugging requires deeper conceptual understanding and reasoning about code execution, it is best supported by understanding and sense-making learning processes. An effective instructional method for fostering such learning is prompted self-explanation \citep{chi1994eliciting, hausmann2007explaining}, which encourages learners to articulate their thought processes, improving their ability to diagnose and correct errors.

\section{Introducing Pedagogical Prompt}
\subsection{What is a Pedagogical Prompt?}
In the context of generative AI, a prompt is generally defined as the natural language input or instruction given to an AI model that guides its generated output \citep{liu2023pre,zamfirescu2023johnny}. 
\textit{\textbf{Pedagogical prompt}} extends this definition by intentionally directing the model to act as a tutor and generate responses that facilitate suitable learning activities aligned with the learner's current stage. For example, a pedagogical prompt may instruct the AI to present worked examples, engage the learner with reflective or guiding questions, or scaffold the problem-solving processes, rather than providing final answers. 

\vspace{-1mm}

\subsection{Main components of a pedagogical prompt}
As illustrated at the top of Figure~\ref{fig: example_prompt}, informed by learning sciences and prompt engineering literature, we identified two key categories that result in the six essential components of a pedagogical prompt: (1) the \textit{learning context}, consisting of five components (highlighted in yellow in Figure~\ref{fig: example_prompt}), and (2) the \textit{learner-instructed tutoring protocol}, consisting of a single component (highlighted in blue in Figure~\ref{fig: example_prompt}).

\subsubsection{Learning Context Components: A pedagogical prompt needs to be well-specified to clearly reflect the learning context} 

A pedagogical prompt requires careful specification of 5 key components to convey the learning context: AI's role, learner's level, problem context, challenge articulation, and guardrails. Specifying \textbf{\textit{AI's role}} improves the tone, knowledge, and reasoning of the output, ultimately enhancing the quality of the LLM-generated responses in both general \citep{zheng2023helpful} and educational contexts \citep{olea2024evaluating}. By explicitly defining the AI's role (such as tutor in our case), the responses become more tailored to student needs and more effective in conveying relevant information. Prompt-writing novices often struggle to include sufficient \textbf{\textit{problem context}} systematically \citep{zamfirescu2023johnny}. This lack of clarity can result in less accurate or misaligned LLM outputs. Therefore, research suggests that incorporating the precise context of the problem, leading to more relevant and accurate responses from LLMs \citep{koutcheme2023automated}. Taking CS education as an example, providing explicit details about the programming task, debugging challenge, or conceptual gap ensures that the AI can generate responses that better align with the learner's needs. Furthermore, according to the CLEAR framework \citep{lo2023clear}, including the \textbf{\textit{learner’s level}} (e.g., beginner vs advanced learner) and \textbf{\textit{difficulty identification}} (e.g., \textit{``I need help debugging''}) can improve the alignment between AI-generated response and the learner's needs. This approach not only improves the accuracy of AI assistance but also supports a more effective and personalized learning experience. Without such clarity, responses can be too advanced, too simplistic, or misaligned with the intent of the learner. Lastly, \textbf{\textit{guardrails}} (e.g., do not provide the direct solution) serve as a type of essential component and proven to be effective for enforcing constraints and ensuring the quality and appropriateness of AI-generated outputs especially for educational purposes \citep{kazemitabaar2024codeaid,bastani2024generative,syah2025edgeprompt}. With these five components providing sufficient learning context, AI can generate more targeted responses.

\subsubsection{Learner-Instructed Tutoring Protocol Component: A pedagogical prompt contains appropriate learner-instructed tutoring protocols based on learning sciences principles} While AI users may sometimes request direct solutions in prompts, especially to complete tasks efficiently in the workplace, prompts designed for pedagogical purposes have a different focus. A pedagogical prompt is a way for learners to tell AI \textit{how} to teach them in their current learning context. Therefore, a good pedagogical prompt should specify the learner-instructed tutoring protocols (e.g., worked example \citep{atkinson2000learning}, repeated practice \citep{nakata2017does}, self-explanation \citep{vanlehn1992model}) to guide effective communication with AI acting as tutors. Because these are learner-instructed, we refer to these as tutoring protocols. The choice of such a tutoring protocol should be (1) aligned with the student's current learning context and (2) supported by learning science frameworks (e.g., the KLI framework \citep{koedinger2012knowledge} in this work).

\begin{figure*}[ht]
    \centering
    \includegraphics[width=1\textwidth]{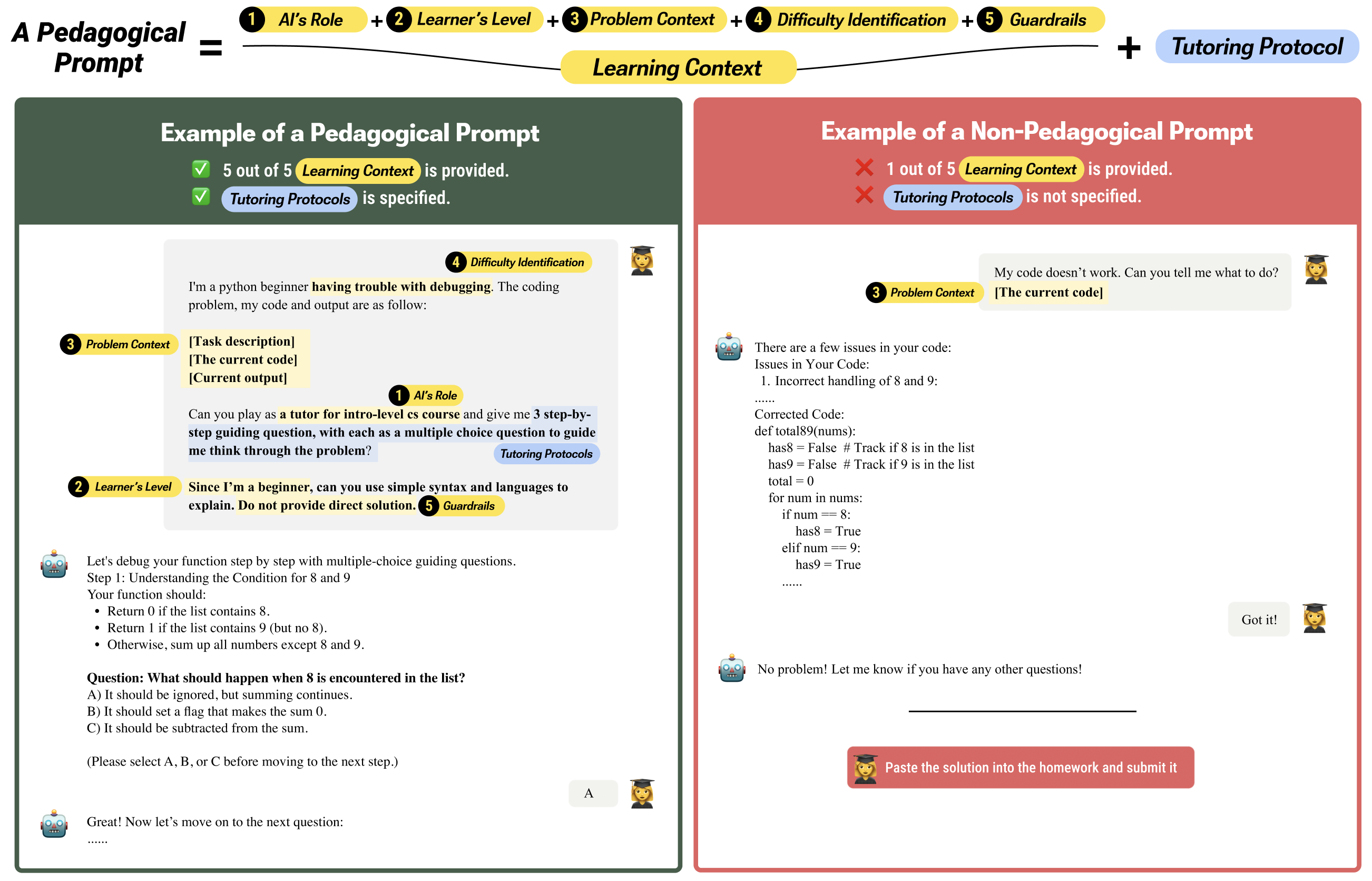}
    \caption{Six components, example and counter-example of a pedagogical prompt. \textbf{Top:} The six essential components of a pedagogical prompt, organized into \textit{Learning Context} (yellow) and \textit{Tutoring Protocol} (blue). \textbf{Left:} A complete example of a well-formed pedagogical prompt in a computing education context, including all five learning context components and a specified tutoring protocol component. \textbf{Right:} A counterexample with minimal context and no tutoring protocol, resulting in a less effective learning experience.}
    \label{fig: example_prompt}
\end{figure*}

\subsection{Example and counterexample implementations of a pedagogical prompt}
Overall, the concept of pedagogical prompts is broadly applicable to different domains or domain-independent. As an initial use case, in this work, we chose to implement the concept of pedagogical prompts within the context of early undergraduate CS education. Figure~\ref{fig: example_prompt} presents both an example (left, green block) and a counterexample (right, red block) of a pedagogical prompt.

In the pedagogical prompt example, the learner provides a rich \textit{Learning Context} by including all five components: AI’s role, learner’s level, problem context, difficulty identification, and guardrails. The learner also specifies a \textit{Tutoring Protocol}, requesting step-by-step multiple-choice questions to guide debugging in an interactive manner. As a result, the AI chatbot follows this protocol by generating guiding questions, pausing for the learner's response, providing feedback, and continuing step-by-step to support an interactive learning process.

In contrast, the counterexample (a non-pedagogical prompt) provides minimal context, only one of five components of the learning context, and does not include a tutoring protocol. Consequently, the AI responds with a direct solution, which the learner passively receives and copies without engaging in the problem-solving process, missing a valuable learning opportunity.

\section{Formative Study: Prove the Value and Understand Instructor Needs for PROOF-OF-CONCEPT Design}
Following the human-centered design method \citep{cooley2000human}, it is important to involve key stakeholders when conducting a proof-of-concept design \citep{stickdorn2012service,abras2004user,sanders2008co}. Therefore, based on the pedagogical prompt concept, we followed the process outlined by \citet{cunningham2021avoiding} to develop proof-of-concept materials through a formative study with targeted instructors. In specific, we distributed a survey to first assess instructors’ willingness to teach pedagogical prompt in early undergraduate CS classrooms. Then we asked about their preferred instructional delivery methods. We distributed it through CS education mailing lists during November 2024 using snowball sampling \citep{parker2019snowball}. This particular sampling method was selected because our target was a specific group of people (instructors teaching early CS courses in higher education), which made random sampling difficult to achieve. This section outlines the survey results and the rationales for our proof-of-concept design.

\subsection{Demographic}
\begin{figure*}[ht]
    \centering
    \includegraphics[width=0.9\linewidth]{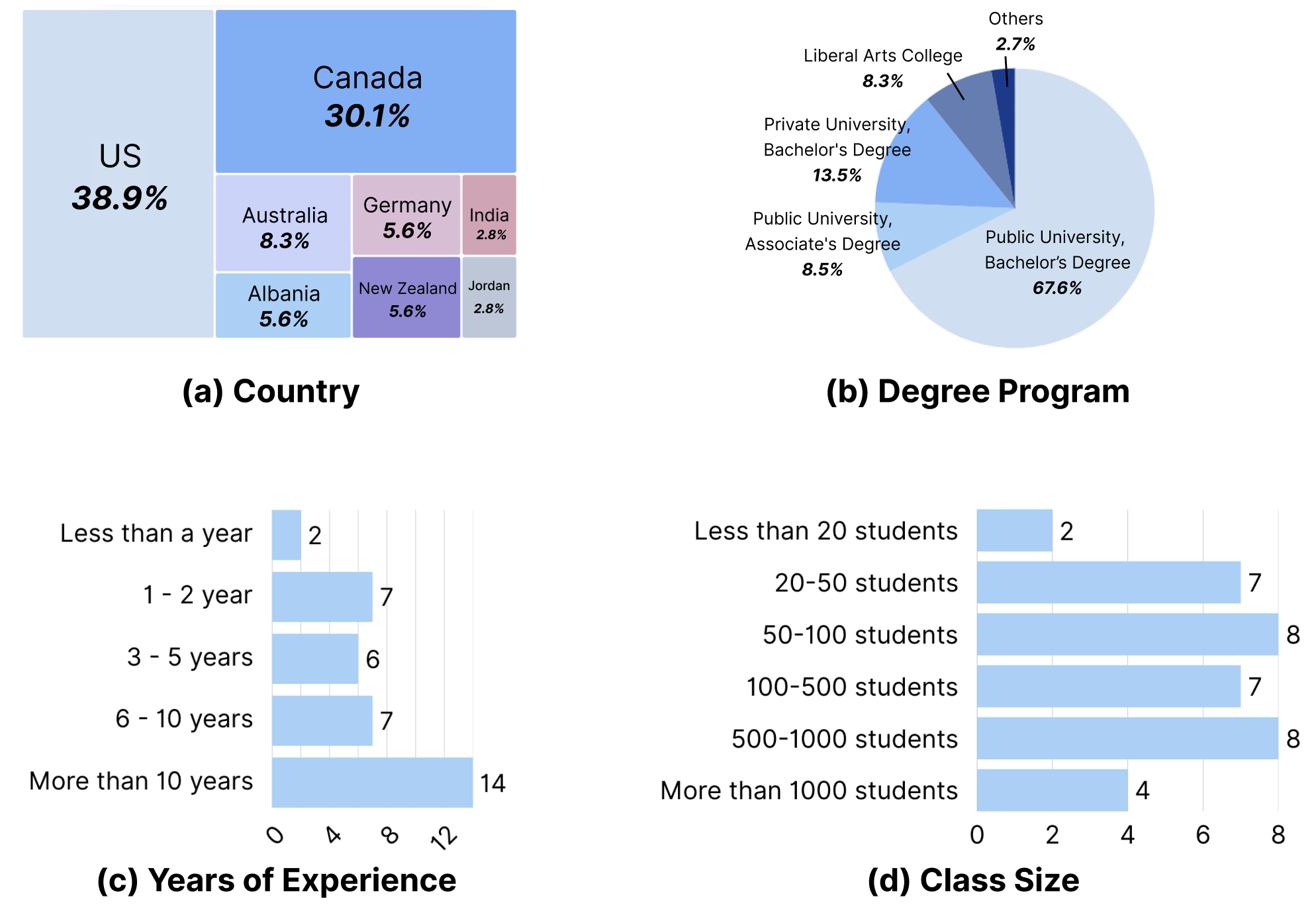}
    \caption{Demographic information of 36 responded instructors: (a) countries; (b) types of degree program their institutions provide; (c) years of teaching experience in early CS courses; (d) common class size.}
    \label{fig: instructor_demographic}
\end{figure*}

We received 36 valid responses from instructors teaching early-stage CS courses in higher education (Figure \ref{fig: instructor_demographic}). Participating instructors come from 8 \textbf{\textit{countries}}, with the majority (69\%) from North America, 14\% from Australia and the rest from Asia and Europe. Regarding \textit{\textbf{degree programs}}, most instructors (69\%) teach at public universities offering bachelor's degree programs. The remainder work in private universities with bachelor’s degree programs (n=5), public universities offering associate degrees (n=3), liberal arts colleges (n=2), and other institutions (n=1). More than half of the instructors have more than five years of \textit{\textbf{experience}} teaching early-stage CS courses (CS1/CS2 courses), and their \textbf{\textit{class sizes}} vary widely, ranging from fewer than 20 students to more than 1000.

\subsection{Confirm the need to teach pedagogical prompting}
More than 90\% of instructors have implemented AI usage policies targeting popular tools like ChatGPT and Copilot, and nearly 80\% of instructors allow different levels of AI usage. Regardless of the varying levels of LLM usage permitted, instructors reported their primary concerns as follows: students relying on AI tools without truly learning the material (50\%), blindly copying solutions (19\%), and engaging in academic dishonesty or cheating (14\%). Furthermore, only 5 instructors (14\%) consider learning pedagogical prompting to be not at all important to novices, and only 1 instructor believes that no instructional time should be allocated to it. In contrast, the remaining instructors reported its varying degrees of importance and expressed their willingness to devote more than one hour of instructional time to teaching this prompting technique. 

These results reinforce the findings from previous literature, highlighting the widespread use of AI in early-stage CS classes alongside growing concerns about its misuse (e.g., asking for direct solutions, cheating, and over-reliance). Lastly, the survey result also reflects instructors’ belief that learning pedagogical prompting is important for all students. This applies regardless of students’ major or their level of knowledge in Computer Science, as more than 50\% of instructors rated the importance higher or equal to 3 on a 5-point scale. Together, these findings underscore the urgent need for teaching pedagogical prompting strategies that promote the effective, learning-oriented use of AI for CS learners regardless of their programming levels and majors.

\subsection{Design decisions for the proof-of-concept in early undergraduate CS education}
More than half (53\%) of the instructors reported that they have taught or know others who have taught students how to use LLMs to learn computer science in early-stage undergraduate CS courses. From their input, we made three key design decisions for the development of our proof-of-concept pedagogical prompting approach in early undergraduate CS education.

\subsubsection{Design-Decision-1: Instructional Method — Delivering the instruction through an interactive learning system} While there are nuanced differences between a system scaffolding prompt selection (\textit{M=3.30} out of 5) versus a system scaffolding prompt writing (\textit{M=3.24}), both are rated higher than more passive instructional methods such as demo videos (\textit{M=2.69}), spreadsheets (\textit{M=2.18}), practice questions (\textit{M=2.61}), or other methods (\textit{M=1.86}). This suggests a clear consensus among instructors for engaging students through active, system-based learning experiences. Therefore, we decided to prioritize implementing a \textbf{\textit{scaffolded prompt construction system}} over instructions through other media.

\subsubsection{Design-Decision-2: System Interaction — Including both prompt selection and prompt writing in the learning intervention} As prompt selection (learners will build a pedagogical prompt by choosing essential components with step-by-step guidance) and prompt writing (learners will construct the prompt by writing each component independently with guidance) are instructors' top-2 choices, differing by only a single vote, we decided to incorporate both interaction types by implementing a \textit{\textbf{select-then-write}} interaction flow for two key reasons. First, the selection step is structured to be a guided discovery experience \citep{bruner1961act}, where we guide learners to focus on the components that are essential for the construction of a pedagogical prompt. This scaffolding, compared to directly having students write the prompt (which can be seen as discovery learning with less guidance), guided discovery reduces cognitive load \citep{sweller2011cognitive, kirschner2010minimal}, increases learning efficiency for such a short intervention time (1 hour), and facilitates deeper engagement \citep{martin2019load}.

However, according to traditional learning theories on learners' engagement (e.g., ICAP framework \citep{chi2014icap}), selecting only is not enough. Thus we included writing as the next-step for the purpose of higher mental engagement on the verbalization process. According to ICAP framework, selecting is classified as an active learning activity, as it involves moving and clicking through the learning materials. However, writing prompts are considered a constructive learning activity that requires students to generate and articulate new ideas. Constructive learning, as outlined in the ICAP framework, leads to higher cognitive engagement and is therefore likely to result in deeper learning and better outcomes for students. Therefore, we implemented such \textit{\textbf{Select-then-Write}} process for the prompt iterating process.

\subsubsection{Design-Decision-3: Length of the Training — Around 1-2 hours} The majority of instructors (69.4\%) indicated a preference for allocating 1–2 hours of instructional time to pedagogical prompting training. Thus, in contrast to other LLM-based tutoring systems that aim to keep learners engaged for extended periods, our system is designed to provide a concise, \textbf{\textit{1–2 hour deliberate practice experience}}.

\section{Learning Intervention: An Interactive System with Scenario-Based Instruction}
This learning intervention is built on an interactive system embedded with scenario-based learning materials. In this section, we describe the technical architecture of the system, the scenario-based instructional materials integrated in the system, and how learners engage with the system to develop pedagogical prompting skills through structured, step-by-step interactions.

\subsection{System Architecture and Technical Details}
The system is developed using JavaScript and features a client-server architecture that facilitates response storage, feedback collection, and communication with OpenAI's API (GPT-4o) for automated prompt validation. The server-side implementation utilizes Node.js through Next.js, which provides REST API endpoints for client-server communication. A custom interaction logger records detailed user interactions, including prompt construction attempts and generated responses. The client-side is built using the \textit{React} framework through Next.js, employing Material-UI for modern, responsive user interfaces.

\subsection{Materials for Scenario-Based Instruction}
We purposefully used scenario-based instruction to reinforce meaningful learning, as it has been shown to be effective when activities are situated in an authentic context \citep{hursen2017investigating}. Since the focus of this study is pedagogical prompt construction, we implemented scenario-based learning centered on help-seeking situations to keep students focused on the pedagogical prompt creation rather than solving the programming problem.

\begin{figure*}[ht]
    \centering
    \includegraphics[width=1\linewidth]{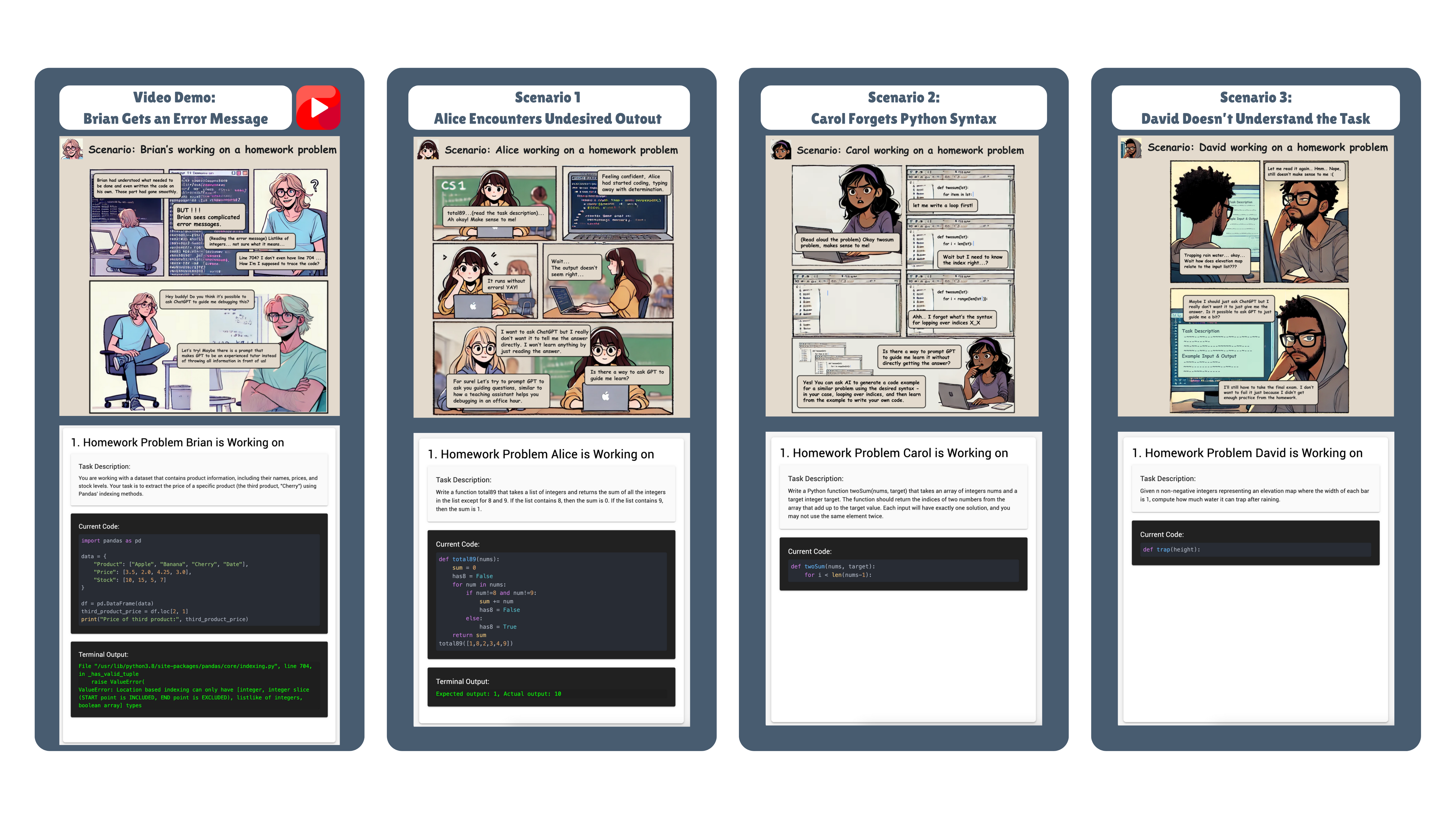}
    \caption{Example scenarios designed for this study, each corresponding to a specific pedagogical prompting learning task. In the user study, Brian’s scenario is presented as the demo video, while the remaining three scenarios are shown to students during the learning session in randomized order.}
    \label{fig: example scenarios}
\end{figure*}

This scenario-based instruction was designed following the five key principles defined in the work from \citet{seren2018scenario}. We selected four common yet unique help-seeking scenarios in early undergraduate CS learning \citep{hou2022using,price2017factors}. In each scenario, the main fictional character (namely, Alice, Brian, Carol, and David; as shown in Figure \ref{fig: example scenarios}) is facing one type of struggle during solving short programming tasks. Students are assigned the role of a peer helping a classmate (the main character). Students are required to thoroughly comprehend the scenarios and construct pedagogical prompts to guide the characters to use AI to help their learning. We gave students the autonomy to explore different prompt element combinations when creating a prompt. We also applied comic-book style narratives to create immersive experience. 

\begin{figure*}[]
    \centering
    \begin{minipage}[b]{0.49\linewidth}
        \centering
        \includegraphics[width=\linewidth]{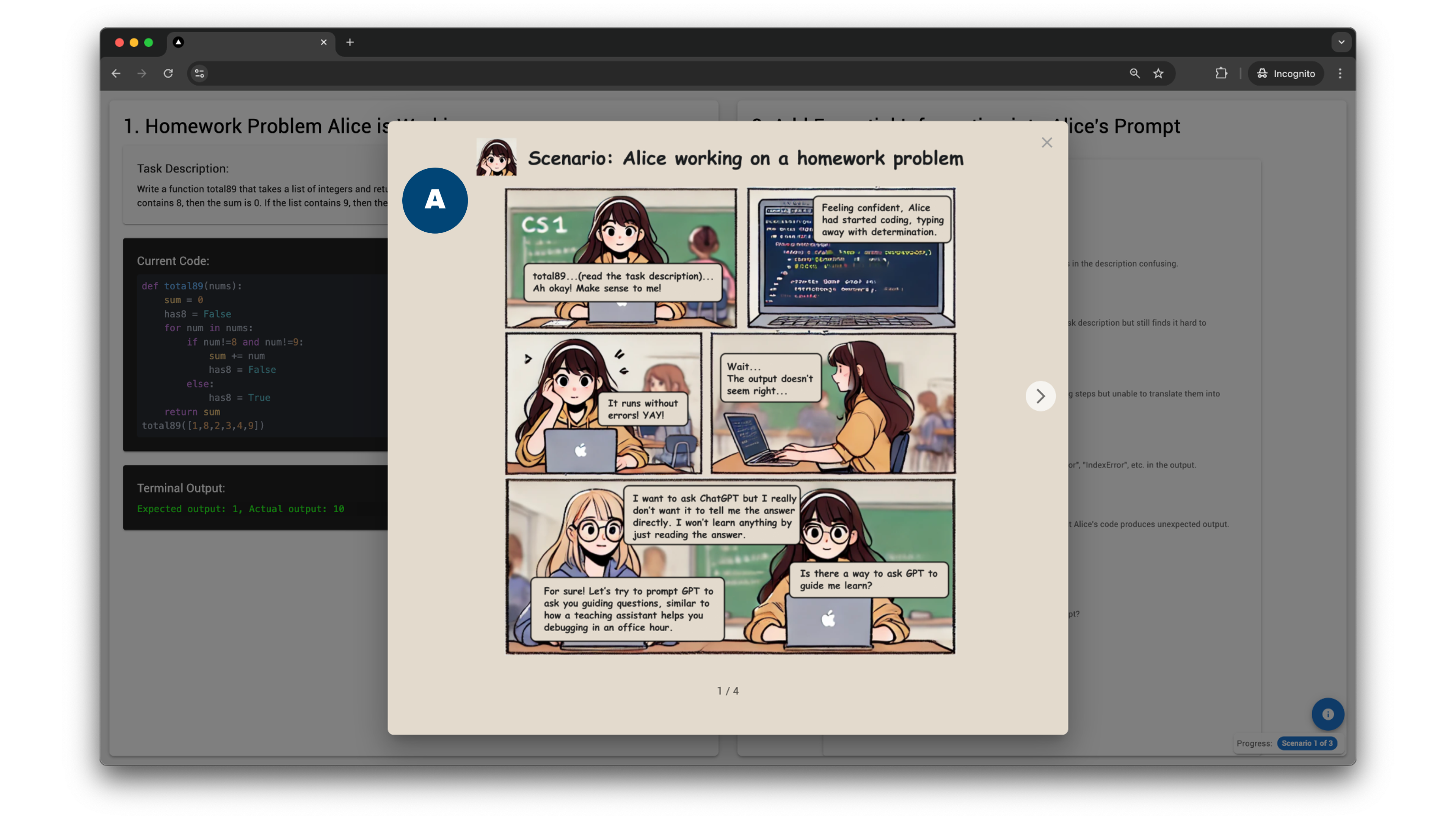}
    \end{minipage}
    \begin{minipage}[b]{0.49\linewidth}
        \centering
        \includegraphics[width=\linewidth]{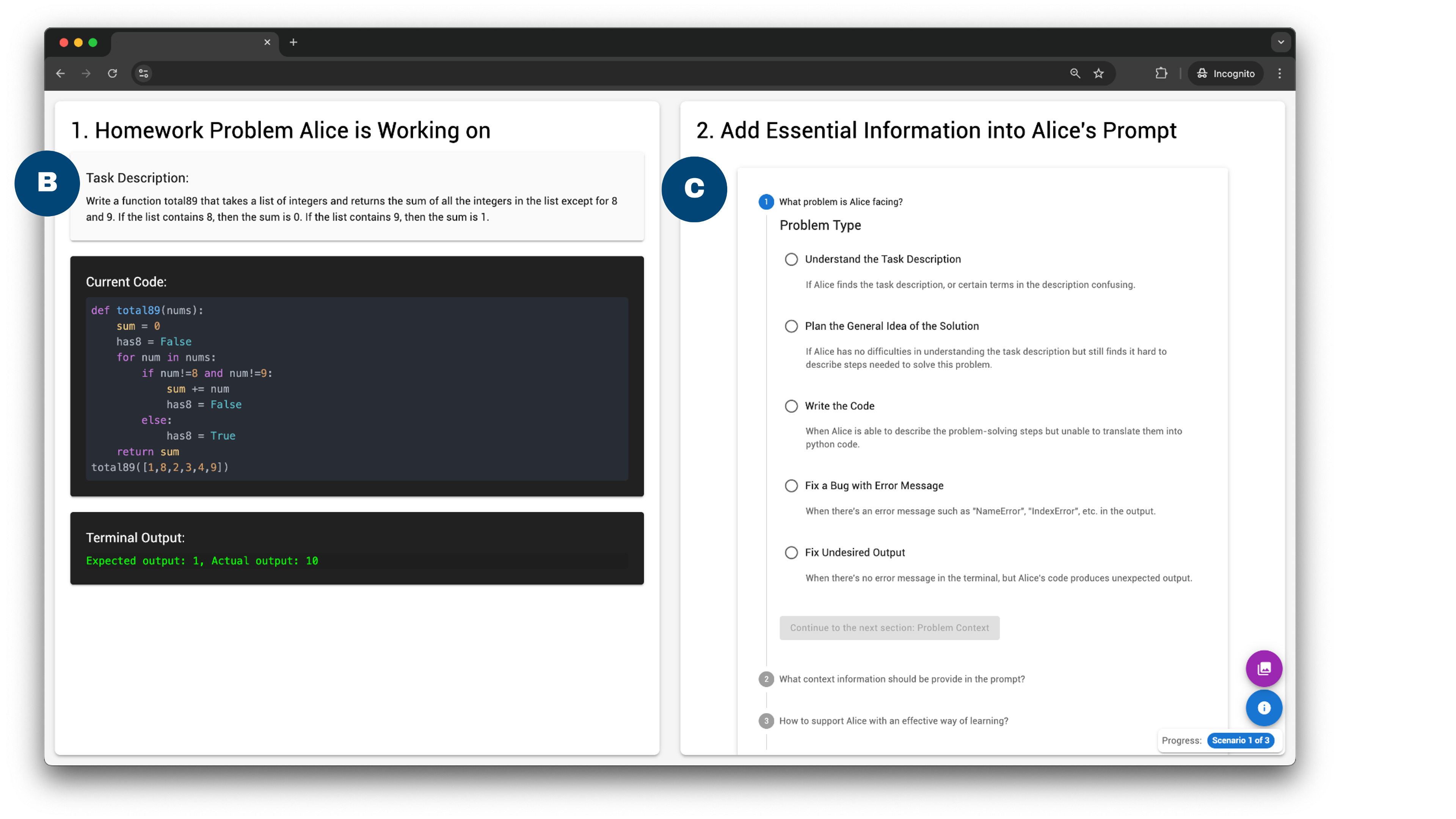}
    \end{minipage}
    \caption{\textbf{(A)} Student first views scene-setting comics for each programming problem (currently shown Alice), then the student interacts with the main learning panel, which consist of \textbf{(B)} left panel: a scenario snapshot including the programming task, the main character’s current code, and its output, and \textbf{(C)} right panel: a step-by-step prompt builder.}
    \vspace{-5mm}
    \label{fig: interface_combined}
    \hfill    
\end{figure*}

When learners begin the experience with our system, they are first presented with a scenario-based comic\footnote{The scenario and content were designed by the authors. Visual elements were generated by OpenAI DALL·E3 based on author-written prompts and subsequently edited by the authors for final use.}. that visually introduces the context of the prompt construction activity (Figure~\ref{fig: interface_combined}-A). The comic sets up a scenario in which learners assist fictional peers in solving programming challenges by crafting effective prompts for ChatGPT. Once students have a clear idea about the scenario and their tasks, they proceed to the learning panel, with the coding interface snapshot (including programming task description, fiction character's current code and current output) of the given scenario on the left panel (Figure~\ref{fig: interface_combined}-B) and a step-by-step prompt builder on the right panel (Figure~\ref{fig: interface_combined}-C). 

\subsection{A Step-by-Step Pedagogical Prompt Builder}
\begin{figure*}[ht]
  \centering
  \begin{subfigure}[b]{0.3\textwidth}
    \centering
    \includegraphics[width=\textwidth]{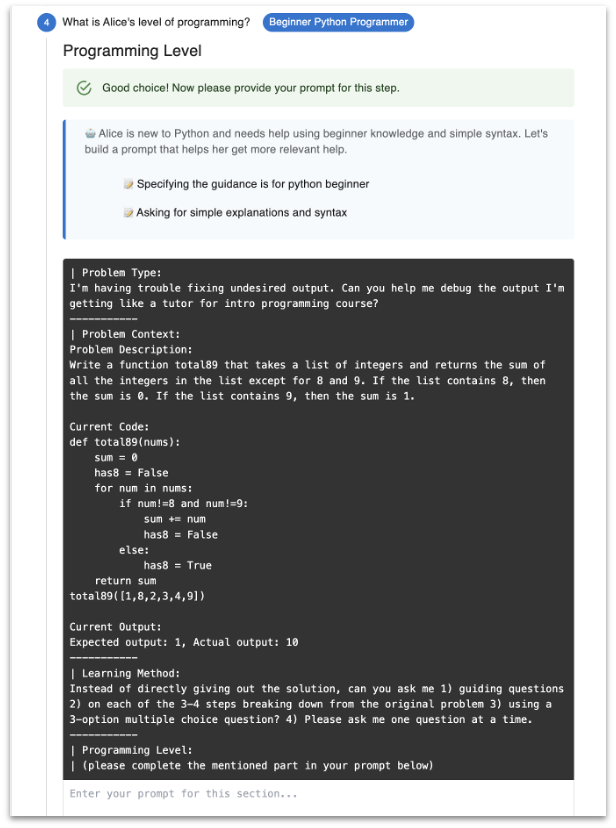}
  \end{subfigure}
  \hfill
  \begin{subfigure}[b]{0.69\textwidth}
    \centering
    \includegraphics[width=\textwidth]{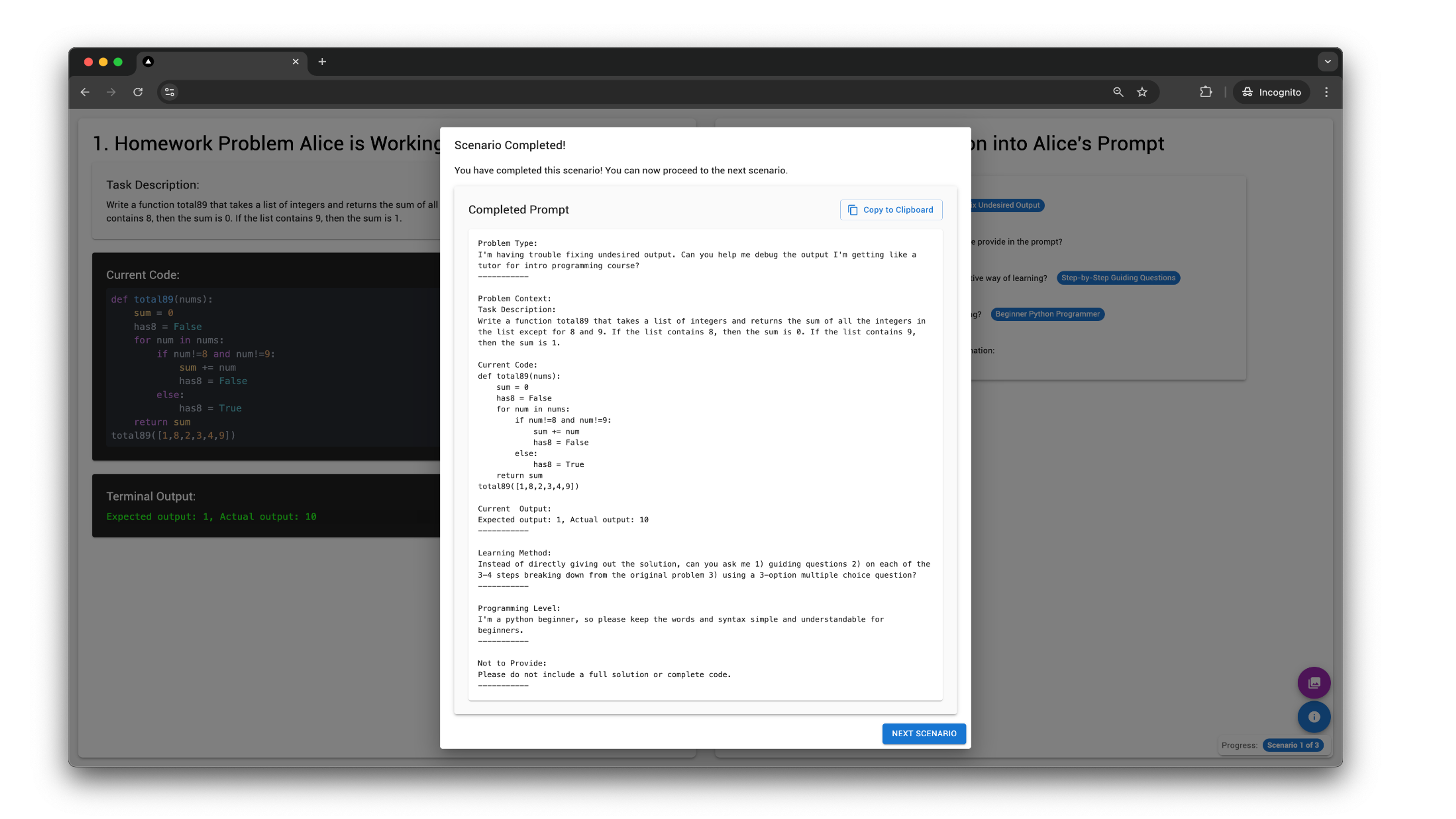}
  \end{subfigure}
  \caption{\textbf{Left}: Previous prompt segments. Students can view the segments they wrote in previous steps above the current input field. \textbf{Right}: Full prompt review. After completing the current scenario, students can view the full prompt built from prior steps.}
  \label{fig: prompt_interface}
\end{figure*}

\begin{figure*}[ht]
    \centering
    \includegraphics[width=1\linewidth]{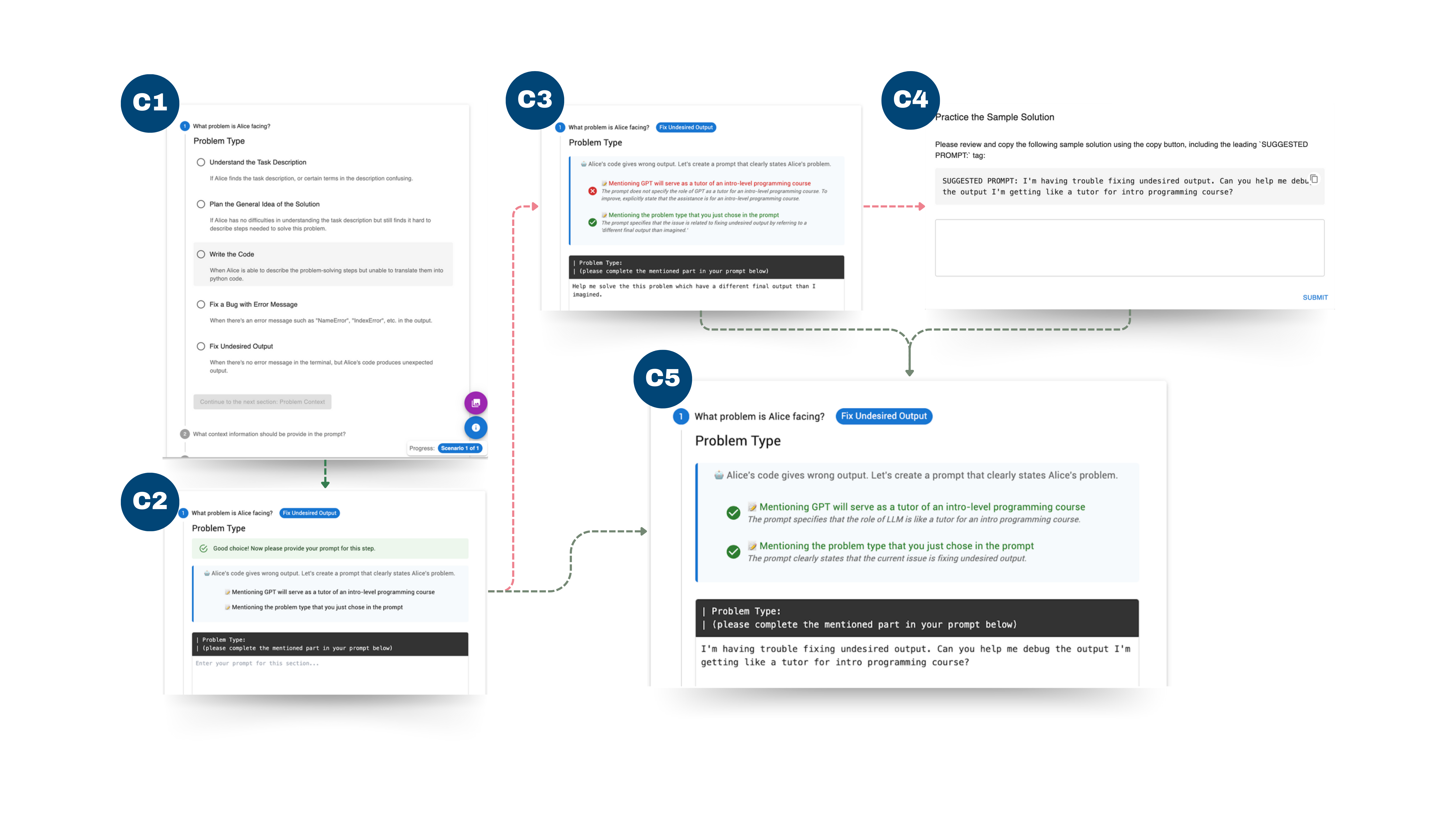}
    \caption{Example of completing Problem Identification step in the prompt builder workflow. The student first (C1) considers and selects the most accurate option within the given list, and then reaches the (C2) prompt writing task, in which they are required to complete the current snippet of the prompt given their previous selection. When the student has drafted a prompt, (C3) an LLM-based feedback system will provide personalized feedback and provide the student with actionable suggestions. If the student still has difficulties passing the prompt validator, (C4) a sample solution will be shown to the student, and they have a chance to review and paste it into the answer box. (C5) When a student have successfully completed the prompt writing task for the current step, they will be directed to the next step.}
    \label{fig: prompt building validation}
\end{figure*}

To effectively scaffold students in creating pedagogical prompts, we divided the prompt construction process into six steps based on the six main components of a pedagogical prompt described in \textit{Section 3}, according to the segmenting principle to reduce unnecessary mental effort \citep{mayer2005principles}. For the \textit{Problem Context} step, the builder lists key information that students must explicitly incorporate into their final prompt and presents it as a short-answer question. Other steps begin with a targeted multiple choice question (Figure \ref{fig: prompt building validation}-C1). After selecting or reading the corresponding step description, participants construct and submit a draft segment of the prompt (Figure \ref{fig: prompt building validation}-C2). When writing a prompt segment, students see their previously completed segments displayed before the current text input area as contextual reminders (Figure \ref{fig: prompt_interface}-Left). Students then submit their prompt segment to receive immediate, AI-generated feedback (Figure \ref{fig: prompt building validation}-C3). We defined specific acceptance criteria for each step of the prompt-building process, and employed OpenAI's \texttt{GPT-4o} (specifically \texttt{gpt-4o-2024-08-06}) to provide personalized, elaborated actionable feedback, highlighting any omissions or areas for improvement. If the student cannot pass all the criteria after three attempts, a sample solution for this step will be provided (Figure \ref{fig: prompt building validation}-C4) to help the student finish the current prompting step (Figure \ref{fig: prompt building validation}-C5). Upon finishing constructing a prompt (Figure \ref{fig: prompt_interface}-Right), students are encouraged to use this prompt to interact with AI, such as putting it into their Chatbot of choice (e.g., ChatGPT, Claude) and observing the AI response.

\vspace{-1mm}
\section{User Study: Understand Student Learning and Perceptions}

To understand the effectiveness of this learning intervention, we conducted a evaluation study with 22 students with IRB (Institutional Review Boards) approval. The RQs are:
\begin{itemize}
    \item RQ1: Does the learning intervention improve students' ability to write pedagogical prompts from pre-test to post-test?
    \item RQ2: Do students' perceptions of their ability to effectively use LLM tools for learning change after this learning intervention?
    \item RQ3: What are students' perceptions of using pedagogical prompts in their future learning?
    \item RQ4: What are students’ perceptions of the usability of this learning intervention?
\end{itemize}

\subsection{Participants \& Study Procedure}
As the current learning intervention grounded in the context of early undergraduate CS education, we recruited students with the targeted programming skill levels. We sent out a recruitment message to a large-scale CS1 class in a public university in northern American. All potential participants scheduled a time with a researcher and the study were all conducted and recorded in Zoom. 

\begin{figure*}[ht]
    \centering
    \includegraphics[width=\linewidth]{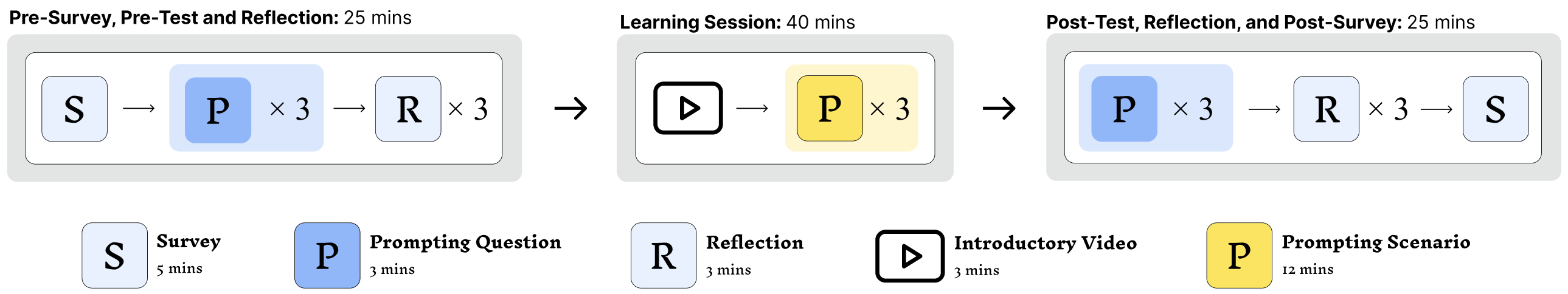}
    \caption{User study procedure. \textbf{Before the learning session}: participants spent 25 minutes to complete a survey, 3 prompting questions in the pre-test, and reflect on the previous 3 prompts; \textbf{learning session}: participants began with watching a 2-3 minutes demo video of the system, and them spent 30-40 minutes to write pedagogical prompts for 3 different scenarios with scaffolding; \textbf{after the learning session}: participants completed 3 prompting questions in the post-test, reflected on the previous 3 prompts, and completed a post-survey in order.}
    \label{fig: experiment_procedure}
\end{figure*}

\begin{figure*}[ht]
    \centering
    \includegraphics[width=0.85\linewidth]{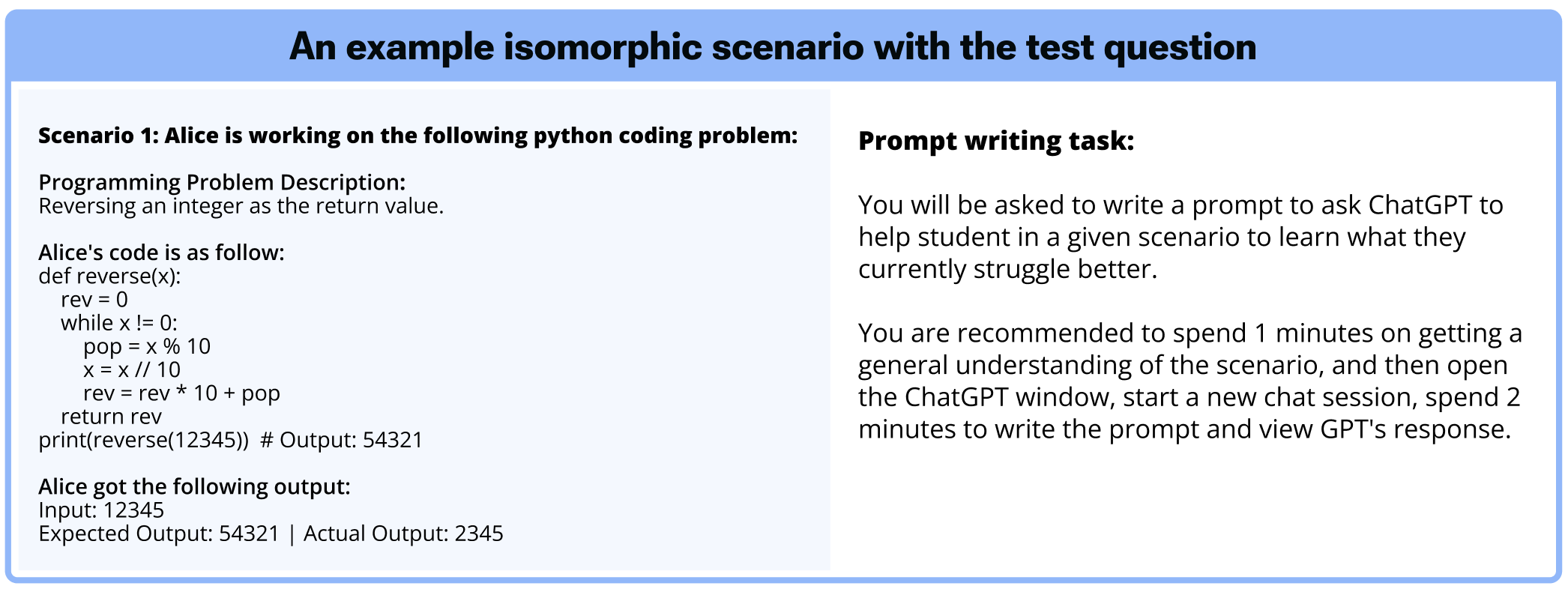}
    \caption{An example isomorphic scenario in pre-test with its corresponding prompt-writing task}
    \label{fig: test_example}
\end{figure*}

At the beginning of each session, participants first completed a consent with a pre-screening survey that included questions about their programming learning level, self-reported familiarity with AI and generative AI tools. Then they got to a pre-test that included three scenario-based prompt writing tasks (Figure \ref{fig: test_example}) with a self-selected commercial AI chatbot, with ChatGPT as the default option. The scenarios were isomorphic across the pre-test, post-test, and learning session. After finishing all three tasks, participants provided reflections on prompts in their responses to help researchers gain a deeper understanding into their thought processes. The learning session started with a detailed on-boarding demo video. After that, participants completed three scenario-based learning tasks using our system in randomized order to decrease the ordering effect. Then they completed a post-test with isomorphic questions as the pre-test, followed by the reflection session. After the reflection, participants completed a post-survey containing 5-point Likert questions and open-ended feedback questions on their learning experiences and interaction with the system. All sections were not strictly time-limited, and participants were allowed to learn and finish the tests at their own pace. Each participant received a \$30 per hour gift card in local currency after the study. The detailed procedure can be found in Figure \ref{fig: experiment_procedure}.

\vspace{-2mm}

\subsection{Data Analysis}
To answer \textbf{RQ1}, two graders manually graded student written prompts in the pre-test and post-test using the rubric extracted from the detailed definition (Section 5.3.1) of a high-quality pedagogical prompt. Specifically, two graders independently graded 15\% of the data, half randomly selected from the pre-test and half from the post-test. The graders reached high agreement in all fields (\textit{Cohen Kappa=0.72} - high agreement in inter-rater reliability \citep{mchugh2012interrater}) and further refined the grading rubric. With the refined rubric, one grader graded the remaining data. For \textbf{RQ2}, \textbf{RQ3} and \textbf{RQ4}, we first summarized and reported students' quantitative survey results, then uncovered the details based on their open-ended answers.

\vspace{-3mm}
\section{User Study Results}

\subsection{RQ1: Students' ability to write pedagogical prompts significantly improved after the learning intervention.}

For each component of a pedagogical prompt, we calculated students' learning gain using \textit{post-test - pre-test}. Then we applied Wilcoxon signed-rank tests \citep{gehan1965generalized} because student test data is not normally distributed. Results show students' (\textit{N=22}) significant learning gains from pre-test to post-test for \textbf{ALL} 6 intended components (\textbf{\textit{p} $<$ .001} (see Table \ref{tab:learning_gains}), indicating the initial success of this learning intervention in teaching pedagogical prompt.

\begin{figure*}[ht]
    \centering
    \includegraphics[width=0.70\linewidth]{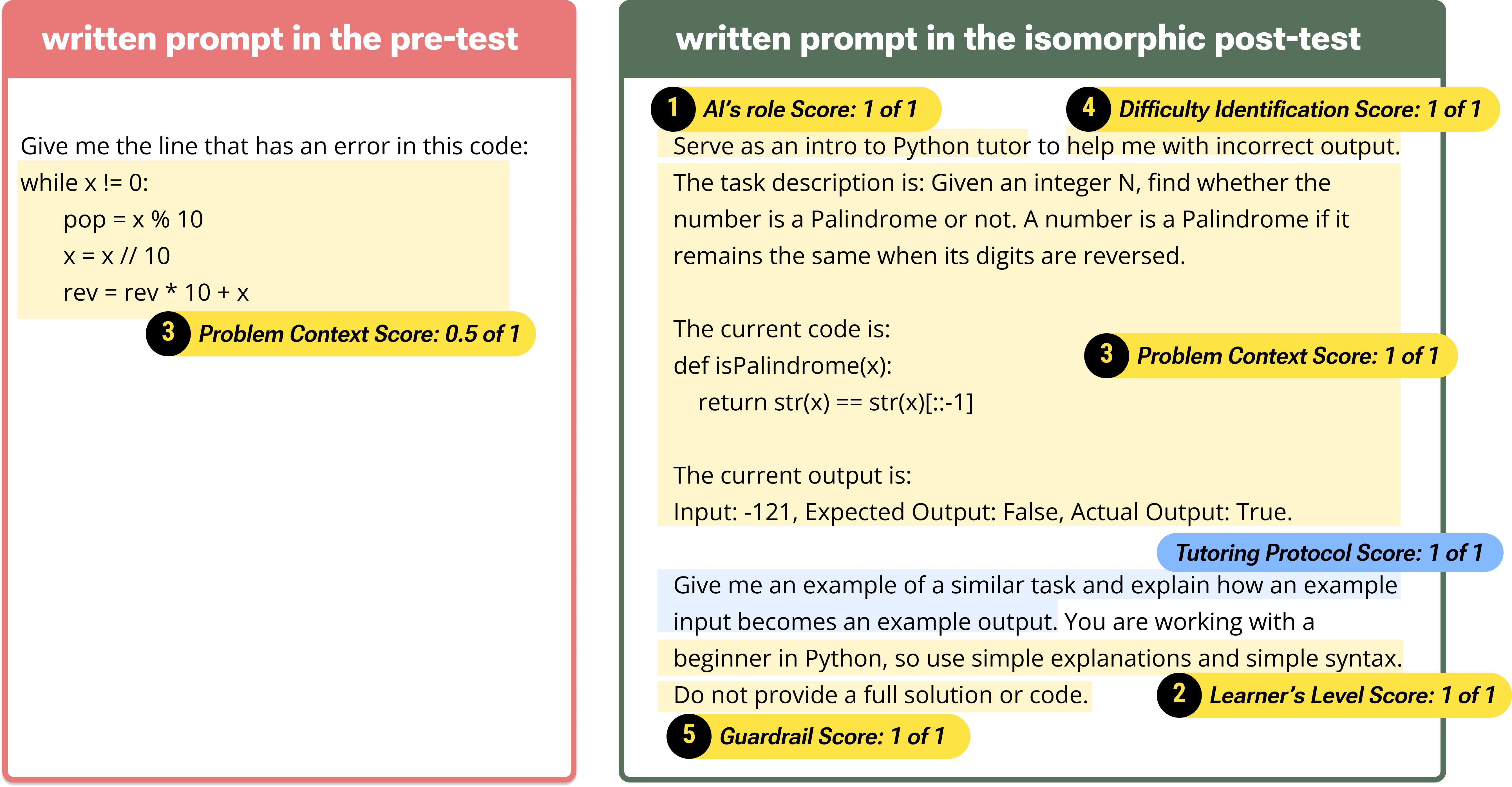}
    \caption{P5’s written prompt to the pre-test question and its corresponding isomorphic post-test question.}
    \label{fig: P5}
\end{figure*}

For example, the complete prompts submitted by participant \textit{P5} in the pre-test and post-test are shown in Figure \ref{fig: P5}. In the pre-test question, \textit{P5} only partially included 1 of the 6 pedagogical prompt components: \textit{Problem Context} (Figure \ref{fig: P5}-Left). After the learning session, when answering the isomorphic \textit{post-test-question\#1}, \textit{P5} came up with a prompt that clearly covered the six required components. The \textit{Tutoring Protocol} was well-articulated as a worked example; sufficient \textit{Problem Context} was provided; \textit{Difficulty Identification} (e.g., \textit{help me with incorrect output}), \textit{AI's Role} (e.g., \textit{as a intro-level programming course tutor}), \textit{Learner's Level} (e.g., \textit{I'm a beginner, so please use beginner-friendly syntax and languages}), and \textit{Guardrails} were all specified in the final written prompt (Figure \ref{fig: P5}-Right).

\begin{table*}[ht]
\centering
\begin{threeparttable}
\caption{Student test performance and learning gain for six pedagogical prompt components}
\label{tab:learning_gains}
\scriptsize
\begin{tabular}{lcccccccccccc}
\toprule
\textbf{} & \multicolumn{3}{c}{\textbf{Pre-Test}} & \multicolumn{3}{c}{\textbf{Post-Test}} & \multicolumn{3}{c}{\textbf{Learning Gain}} & \multicolumn{3}{c}{\textbf{Statistics}}\\
\cmidrule(lr){2-4} \cmidrule(lr){5-7} \cmidrule(lr){8-10} \cmidrule(lr){11-13}
\textbf{Criteria} & \textbf{\textit{M}} & \textbf{Median} & \textbf{IQR} & \textbf{\textit{M}} & \textbf{Median} & \textbf{IQR} & \textbf{\textit{M}} & \textbf{Median} & \textbf{IQR} & \textbf{z} & \textbf{p-value} & \textbf{Effect Size} \\
\midrule
\textit{Tutoring Protocols}         & 0.44 & 0.0 & 1.0 & 0.83 & 1.0 & 0.0 & 0.39 & 0.5 & 1.0 & 9.46 & $<$.001\textsuperscript{***} & 0.82 \\
\textit{Difficulty Identification}  & 0.70 & 1.0 & 1.0 & 0.95 & 1.0 & 0.0 & 0.26 & 0.0 & 0.8 & 9.91 &$<$.001\textsuperscript{***} & 0.86 \\
\textit{AI's Role} & 0.00 & 0.0 & 0.0 & 0.62 & 1.0 & 1.0 & 0.62 & 1.0 & 1.0 & 9.91 & $<$.001\textsuperscript{***} & 0.86 \\
\textit{Problem Context}         & 0.77 & 1.0 & 0.5 & 0.94 & 1.0 & 0.0 & 0.17 & 0.0 & 0.5 & 9.63 & $<$.001\textsuperscript{***} & 0.84 \\
\textit{Learner's Level}           & 0.08 & 0.0 & 0.0 & 0.65 & 1.0 & 0.9 & 0.58 & 1.0 & 1.0 & 9.75 & $<$.001\textsuperscript{***} & 0.85 \\
\textit{Guardrails}    & 0.25 & 0.0 & 0.8 & 0.82 & 1.0 & 0.0 & 0.56 & 1.0 & 1.0 & 9.72 & $<$.001\textsuperscript{***} & 0.85 \\
\bottomrule
\end{tabular}
\begin{tablenotes}
\scriptsize
\item \textit{* $p$ $<$ .05, ** $p$ $<$ .01, *** $p$ $<$ .001. Wilcoxon signed-rank test used for non-parametric data; interquartile range (IQR) is reported to describe variability in scores.}

\end{tablenotes}
\end{threeparttable}
\end{table*}
    
\subsection{RQ2: Students perceptions of their ability to effectively use LLM for learning have been corrected and increased after the learning intervention.}

\begin{figure*}
    \centering
    \includegraphics[width=\linewidth]{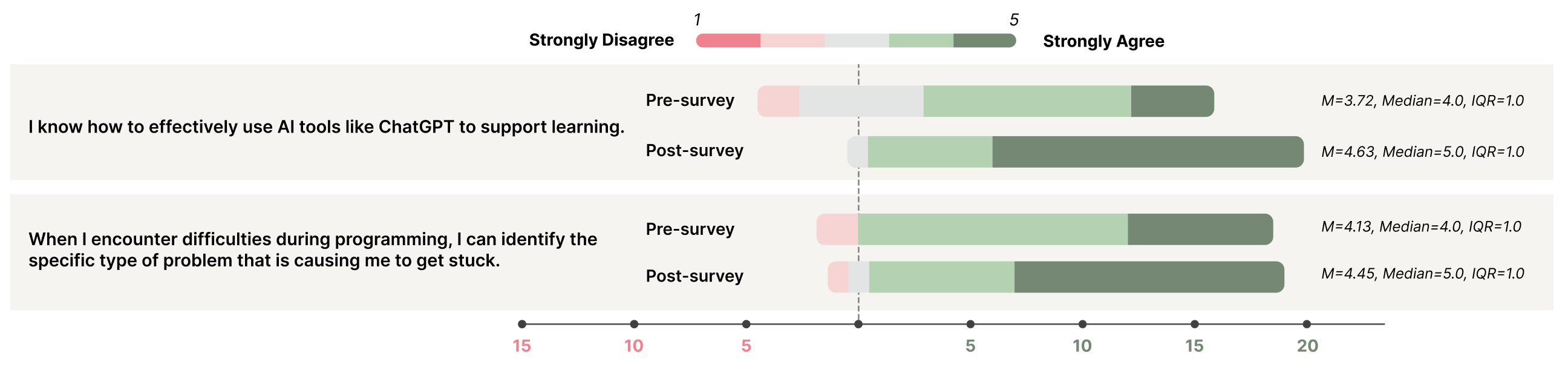}
    \caption{Changes in students’ self-reported ability to use LLM tools for learning before and after the learning intervention}
    \label{fig: paired_likert}
\end{figure*}

Students reported relatively high confidence in their ability to use LLMs effectively for learning in the pre-test (Figure~\ref{fig: paired_likert}, first row; \textit{M} = 3.72 of 5). However, their self-reported confidence did not accurately reflect their actual prompting ability. Specifically, if students had an accurate self-assessment of their skills, a significant positive correlation between confidence and pre-test performance would be expected \citep{baxter2011self}. In contrast, in our result, a Pearson correlation showed a negative but not significant relationship between the perceived confidence score and pre-test performance (\textit{r} = –.15, \textit{p} = .498), suggesting a misalignment between the perception and actual abilities of the learners before the intervention. This pattern aligns with findings from \citet{prather2024widening}, suggesting that students may hold a false sense of confidence in their LLM-based learning, which can potentially lead to conceptual gaps and falling behind in course content. After the learning session, participants’ perceived confidence increased to an average of 4.63 of 5 (Figure \ref{fig: paired_likert}-row 2), reflecting both greater perceived confidence from pre-survey to post-survey (\textit{p} = .001, \textit{z} = 5.38, \textit{effect size} = .81, Wilcoxon signed-rank test) and a significant positive alignment between their perceived ability and actual performance (\textit{r} = .54, \textit{p} = .032, Pearson correlation). Furthermore, as indicated in Figure \ref{fig: paired_likert} (row 3 to row 4), students reported perceived improvements in identifying their specific struggles, although the improvement is not significant (\textit{p} = .071, \textit{z} = 5.26, Wilcoxon signed-rank test).

To further understand the potential factors that might contribute to changes in perceived learning gains, we took a look at students' responses to feedback questions in the post-survey. Students highlighted factors such as increased awareness of LLMs' instructional potential (e.g., P12-\textit{``I now know how to simulate office hours environments with GPT.}), being more self-regulated (e.g., P17-\textit{``I was not giving myself the opportunity to learn...'' }, P20-\textit{``...helps with actually understanding and promotes academic integrity''}), and increased prompting skills. A common aspect mentioned by all participanting students is the clear distinguish between asking for direct solution and learning-oriented use (e.g., P11-\textit{``Now I know how to prompt ChatGPT to act as an entry-level tutor that provides guidance instead of solutions''}).

\begin{figure*}
    \centering
    \includegraphics[width=\linewidth]{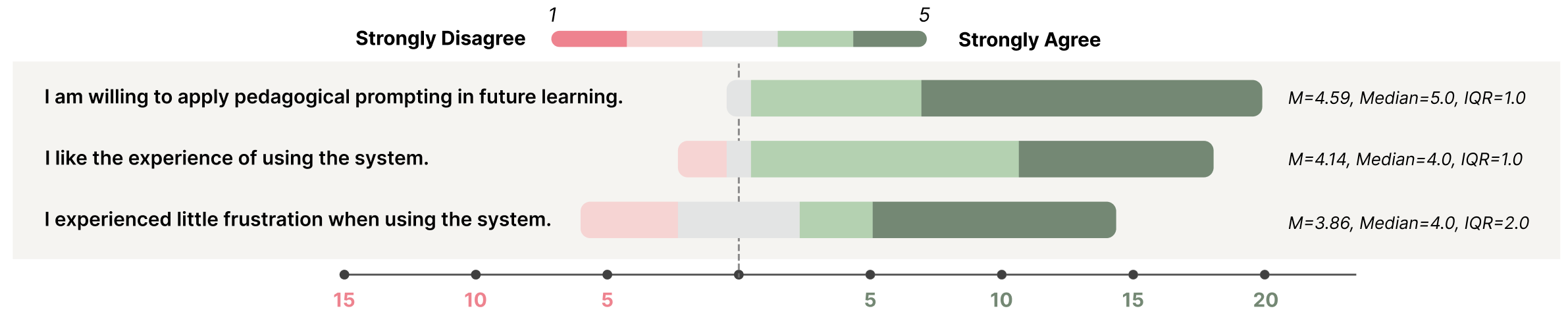}
    \caption{Students’ willingness to use pedagogical prompting in future learning and perceptions of the system experience reported in the post-survey}
    \label{fig: rest_likert}
\end{figure*}

\subsection{RQ3: Most students are extremely willing to apply pedagogical prompts in their future learning.}

After the learning intervention, we also asked participants about their willingness to apply pedagogical prompts into their future learning on a 5-point Likert scale. Around \textbf{72.7\%} of students reported being extremely willing (5 out of 5 in the likert question) to continue using them, while five participants indicated \textit{``somehow agree''} to continue using them, and one participant remained neutral (1st row in Figure \ref{fig: rest_likert}). Students are also asked in the survey to reflect on using generative AI in a pedagogical way compared to just asking the AI to provide the solution. The most common pattern was a consistent acknowledgment among all participants that AI producing direct answers was faster and easier, but also led to minimal understanding. However, writing pedagogical prompts is more challenging and more beneficial, promotes learning, and independent problem-solving. Moreover, 8 students demonstrated strong self-regulation in their reflections. For example, \textit{P15} said \textit{``I would like to use it as it forces me to first understand which step of the overall task I am stuck at e.g., Understanding the Task, Writing the Code, Debugging, etc...''}; \textit{P16} would like to \textit{``definitely try out pedagogical prompting during my studies for my [course id] exam''}. These examples suggest that students became more aware of the types of problem they encounter and more able to recognize when it is appropriate to use pedagogical prompts.

Nevertheless, six students expressed reluctance or occasional unwillingness to use pedagogical prompting in their learning process. Three of them mentioned that the prompt is \textit{``too long''} (\textit{P10,P17,P22}), or believe \textit{``the learning gain is marginal''} (\textit{P17}) and believed that they can still learn from direct solutions (\textit{``I don't mind getting a direct answer and then trying to comb through to figure out how that answer was generated.'', P3})

\subsection{RQ4: Students rated the system highly, despite occasional moments of frustration.}
With a high rating on their \textbf{system using experience} (2nd row in Figure \ref{fig: rest_likert}, \textit{M} = 4.14), when asked about their perceptions of the system that scaffolded their prompt-writing experience, several key benefits emerged. Many students appreciated how the system helped them understand AI's role in learning beyond just providing answers, with \textit{P22} noting, \textit{``It really taught me how AI can be helpful in learning and not just for getting answers,''} and \textit{P23} describing it as an opportunity to \textit{``figure out a better way to use AI for learning.''} Several participants valued the practice in prompt engineering (\textit{P4}) and the structured guidance in writing effective prompts, as \textit{P5} explained, \textit{``It helped me understand different parts of making a good prompt.''} The scaffolding was seen as beneficial, with \textit{P12} mentioning that \textit{``writing the requirements for each part of the prompt made it easier,''} while \textit{P21} highlighted its\textit{ ``step-by-step prompt construction.''} Additionally, the system’s design was well-received for being ``easy to use'' (\textit{P14}), \textit{``straightforward and simple to use''} (\textit{P17}), and \textit{``very user-friendly to navigate''} (\textit{P20}). The feedback component was particularly appreciated, with \textit{P17} stating, \textit{``the system would tell me exactly what I got wrong''} during the prompt-construction process. Some students also enjoyed interactive elements, such as the comics, which \textit{P12} said \textit{``made the task more enjoyable.''}

However, some participants did experience \textbf{frustration}. While more than half reported little to no frustration, four participants indicated experiencing noticeable frustration during the interaction (3rd row in Figure \ref{fig: rest_likert}). A few participants found the system occasionally inaccurate or too rigid in evaluating prompts, as \textit{P21} noted, \textit{``In a few cases, I believe my input met the criteria, but it did not accept it because of a minor difference in wording.''} Others found it \textit{``annoying at times since it sometimes randomly rejected what I wrote''} (\textit{P4}) or \textit{``frustrating at times to reach the correct answer''} (\textit{P14}). The process was also perceived as too lengthy or tedious by some (\textit{P7, P10}), with \textit{P8} commenting that certain prompt requirements, such as \textit{``act as a tutor''} and \textit{``I’m a Python beginner,''} felt redundant. Lastly, \textit{P19} expressed a wish for easier scenario recall, stating, \textit{``I wish that I could re-access the scenarios because I sometimes forgot what was happening.''} These insights highlight both the strengths of the system in fostering structured prompt writing and the areas where improvements could enhance the user experience.

\vspace{-1mm}
\section{Discussion}

Enhancing the pedagogical quality of the human-AI interaction is increasingly critical to supporting meaningful learning, especially as most of the students in our study recognized the value of pedagogical prompts and expressed a strong willingness to use such prompts in the future. This aligns with growing concerns in the literature about the potential harm of misusing AI in educational settings \citep{bastani2024generative}, as well as calls for more intentional and pedagogically grounded applications of AI tools \citep{prather2025beyond}. 

This work focuses on identifying actionable next steps for early CS courses in response to the rapid changes introduced by generative AI (GenAI). It is grounded in the assumption that, in the near term, CS curricula will continue to emphasize foundational programming skills, even as GenAI tools become increasingly integrated into the learning process. As an initial response to this shift, our findings offer insights into both immediate instructional adjustments and longer-term strategies to redesign AI-enhanced learning experiences. We identify three key stakeholders in this evolving landscape: teachers, learners, and AI systems, and discuss the unique insights and opportunities associated with each role.

\vspace{-2mm}

\subsection{Instructors: need for an interactive system that delivers brief instructional activities to teach pedagogical prompting for all CS novices}
The survey results of 36 early CS course instructors confirmed the importance of teaching pedagogical prompting, established its priority within instructional goals, and produced implementation guidelines. In summary, instructors prefer a brief (1–2 hours) deliberate practice session for CS novices, regardless of their major, using an interactive system to practice constructing pedagogical prompts. 

\subsubsection{\textbf{Instructors prefer a brief but focused pedagogical prompting instruction session}}
The medium-level priority and preferred 1-2 hours' instructional duration reveal the status quo of CS education in the era of generative AI (GenAI). On the one hand, instructors believe that the primary goal of early-stage CS courses (CS1/CS2) should still be to go through traditional programming topics \citep{wing2006computational,denny2024computing}. Therefore, most of them did not rank either top priority or wanted to put excessively long instruction hours to teach pedagogical prompting. On the other hand, GenAI is drastically reshaping CS education \citep{prather2025beyond} and education in general \citep{qadir2023engineering}. Recognizing its importance, instructors are willing to allocate portions of their regular teaching time to cover it as a topic. 

\subsubsection{\textbf{Pedagogical prompting should be taught to all students in CS classes, regardless of their major and level of programming.}}
There is a consensus on the targeted student population. Instructors think both CS majors and non-majors should learn pedagogical prompting, regardless of their major or level of programming. Recent studies support providing AI education to this broad audience \citep{shernoff2024computer,walter2024embracing,denny2024computing}. For instance, \citet{walter2024embracing} emphasizes the importance of AI literacy and prompt engineering for all students regardless of their major programs, highlighting their role in fostering critical thinking skills essential in the age of AI \citep{walter2024embracing}. Similarly, \citet{denny2024computing} discuss the transformative impact of generative AI on computing education, underscoring the need for students across disciplines to engage with these tools thoughtfully \citep{denny2024computing}. These findings suggest that equipping all students with the skills to use AI tools pedagogically is crucial. It's not only for students' own academic success, but also for their preparedness in a workforce increasingly influenced by AI technologies. 

This agreement of instructors can also be explained by their shared concerns. Those concerns are more broadly about academic misconduct and responsible use of AI tools, instead of CS majors-specific ones, such as reduced competitiveness in the software engineering job market or inadequate prior knowledge for advanced-level courses. These possible explanations suggest that the rationale for teaching pedagogical prompting lies not in tailoring instruction solely for future CS professionals, but in addressing broader educational needs and ethical challenges that impact all learners navigating AI-enhanced learning environments. 

\subsubsection{\textbf{An interactive instructional system is the most preferred way for delivering the training}}
Regarding the medium of delivering pedagogical prompting instruction, an interactive system is the top choice. This finding demonstrates instructors' preference for a more engaging instructional method. Their preferred ways align well with the ICAP framework \citep{chi2014icap}, which categorizes learning activities by the level of cognitive engagement they promote. Interactive instructional systems encourage students to either construct prompts themselves or actively select and refine them through system-guided scaffolding, both of which fall into the \textit{Constructive} or \textit{Interactive} modes of engagement. These levels are shown to produce deeper learning outcomes because learners generate new knowledge or meaning beyond the information presented. In contrast, alternatives such as demo videos or static spreadsheets primarily foster \textit{Passive} or at best \textit{Active} engagement, where students receive or manipulate information without generating new understanding. Practice questions can also be a viable option; however, compared to an interactive system, they may lack personalized, immediate feedback \citep{shute2008focus}, as well as autonomy, which could induce boredom according to \citet{pekrun2014control}’s Control-Value Theory of Achievement Emotions \citep{pekrun2014control}. 

\subsubsection{\textbf{Selecting vs. Writing: Designing Interactions to Balance Efficiency and Active Skill Development}}
Lastly, although our current system adopts a selecting-then-writing interaction design, we aim to further investigate which interaction method (selecting or writing) optimally balances learning efficacy and student engagement. While we hypothesize that writing, as a \textit{Constructive} activity within the ICAP framework, may lead to deeper learning despite requiring greater cognitive effort (and potentially inducing boredom), some participants reported frustration with this approach. As a next step, we plan to conduct an A/B test in classroom settings to systematically compare the two interaction methods and identify the most effective and engaging design for pedagogical prompting instruction.

\subsection{Learners: Developing Cognitive and Metacognitive Skills to Strengthen Learner Agency in the Age of Generative AI}

\subsubsection{\textbf{Prompting skill is a pressing need for nowadays learners, and scaffolded writing practice is an effective, immediate next-step that can significantly increase learners' prompting ability in using AI for learning}} 
As generative AI becomes an inevitable component of future learning environments, it is crucial to prepare students to engage with it effectively rather than passively or improperly \citep{prather2025beyond}. Yet, learners, especially novices, often lack the strategies for effective AI-based help-seeking \citep{xiao2024preliminary}, and the misuses of AI can negatively impact learning outcomes \citep{bastani2024generative}. Therefore, equipping students with the skills to use AI tools purposefully and critically will be an essential immediate next-step in education. 

Our system provides a solution to achieve this instructional goal in prompting learning in a short time period, as indicated by the promising results of the user study. After learning with our system, both learners' ability and perceived confidence in pedagogical prompting significantly improved from pre-test to post-test. As shown in Table \ref{tab:learning_gains}, participants made the most substantial gains in articulating the \textit{Learning Method} and specifying the \textit{AI’s Role}, which are critical components of effective pedagogical prompting. These improvements suggest that learners became more capable of providing structured, instructional guidance to the AI, thus enhancing the overall quality and clarity of their prompts. Mapping these gains to \citet{nelson1981help}'s five-stage help-seeking model, we can observe strong developments in Stage 3: \textit{Identifying a potential helper}, and Stage 4: \textit{Employing strategies to elicit help}. By learning how to clearly specify the learning method and contextualize the AI’s role, students demonstrated an improved ability to recognize the AI as a valid and appropriate source of help for learning tasks. More importantly, they learned to strategically phrase their prompts in ways that actively engage the AI in a pedagogically useful manner.

This success may be attributed to the system’s ability to stimulate learners’ psychological engagement \citep{stull2007learning} through repeated practice \citep{pedersen2016diffgame} on constructive-level tasks \citep{chi2014icap}. We deliberately designed an interaction-heavy system, recognizing that while many existing works assume a high degree of learner self-regulation, this assumption often does not hold in real-world educational settings, as learners frequently demonstrate unregulated or suboptimal learning behaviors \citep{jansen2020supporting,molenaar2023measuring,xiao2024preliminary}. To address this, we embedded interactions that involve productive struggles \citep{gresalfi2009constructing} and are difficult to "game," encouraging meaningful engagement without prerequisites on learners' self-regulation. Since the effectiveness of the learning experience has been demonstrated, a next-step for future research is to investigate the system's learning efficiency by optimizing learning gains while minimizing potential redundancy and learner frustration.

\subsubsection{\textbf{Meta-cognitive skills can further strengthen students' ability to manage AI-enhanced learning}} Meta-cognitive skills such as self-regulated learning (SRL, the ability to plan, monitor, and reflect on one’s own learning) can be essential in AI-enhanced environments where learners must navigate open-ended tools like generative AI \citep{panadero2017review}. At the emergence of GenAI in 2022, many students used tools like ChatGPT primarily to obtain direct answers, often bypassing meaningful engagement. However, with more than two years of widespread adoption, our study, although based on a small sample size, reveals a reassuring shift on learners' self-regulation: most students now recognize that copying AI-generated solutions does not support learning. Yet, as shown in our pre-test results (Section 7.1), they still struggle to prompt effectively for pedagogical support. This gap highlights a lack of SRL, particularly in the strategic planning and help-seeking phases. Without clear strategies for articulating their learning goals or identifying appropriate instructional guidance, students fail to leverage AI for long-term understanding. Future works that regulate SRL (e.g., by prompting goal-setting, self-explanation, or reflection) can support learners in becoming more intentional and self-aware users of AI, ensuring these tools serve as learning partners rather than shortcut providers \citep{baker2023ai, ainley2022student}.

\subsection{AI Systems: Providing more accessible learning experiences with pedagogical and data-driven features.}
While our intervention demonstrated that students can learn to construct pedagogical prompts, participants’ concerns about the length and cognitive effort involved should not be overlooked. Several learners reported a preference for receiving effective output rather than investing in the skill of prompt construction itself, a tension reflecting the distinction between instrumental use of tools and epistemic engagement in learning activities \citep{rabardel1995hommes, roll2012evaluating}. Without adequate support, such prompting process can introduce extraneous cognitive load \citep{sweller1998cognitive}, potentially undermining learning by overwhelming novice learners' limited working memory resources.

\subsubsection{\textbf{Implementing Pedagogical Features}}
Given that learners are not expected to possess deep knowledge of learning sciences or instructional design, the responsibility for pedagogical alignment should not rest solely on them. Instead, AI systems should increasingly shoulder this burden by embedding pedagogical features directly into the prompting phase. One promising direction is to offer structured prompt templates or pre-configured scaffolds encapsulated pedagogical strategies (e.g., worked examples, reflective questions) that learners can select from with minimal effort \citep{kazemitabaar2024codeaid,xiao2024exploring}. This not only reduces the cognitive barrier to engaging in pedagogical prompting but also supports students with varying levels of self-regulation and AI literacy.

\subsubsection{\textbf{Implementing Data-Driven features}}
Moreover, learners often feel burdened by the need to articulate their knowledge level and learning progress each time they prompt, as noted by participants in Section 7.4. This burden can be reduced if the AI system is capable of inferring such information using data-driven methods. Several GenAI tools already adopt this strategy. For example, QuickTA \citep{kumar2024supporting} uses reinforcement learning and rule-based logic to adapt help levels based on student interaction without requiring explicit learner self-disclosure. However, core inference techniques from traditional intelligent tutoring systems (ITS) remain underutilized in GenAI. For example, Performance Factor Analysis (PFA) \citep{pavlik2009performance} is often used in traditional ITS to predict future performance by weighting past successes and failures per skill. Methods like PFA, proven effective in ITS, can enhance GenAI tutors by automatically personalizing instruction without requiring learners to verbalize their progress repeatedly. Integrating such techniques with LLM-based systems can significantly reduce learners' effort and improve adaptive support.
\section{Limitations and Future Work}

\subsection{Toward Longitudinal Evaluation of Pedagogical Prompting}
As an initial step, this study evaluated the short-term effectiveness of learners’ prompting skill development. The longer-term impacts of pedagogical prompting on programming learning were primarily based on premises from educational theory and prior literature (e.g., the established effectiveness of worked examples in teaching procedural knowledge). To more rigorously assess the effectiveness of pedagogical prompting, future research should involve large-scale classroom studies that investigate its long-term effects on help-seeking and programming learning outcomes. These studies should also include analysis of usage frequency, interaction logs, and formative and summative assessment scores to offer a more comprehensive understanding of the participation of the learner and the impact of the instructional.

\subsection{More Accurate Measurement on Learners’ Illusion of Competence and Confidence Calibration}

Learners often exhibit an illusion of competence in the pedagogical use of AI (Section 7.2). Our analysis of RQ2 reveals a discrepancy between the actual abilities of the students to use AI pedagogically and their corresponding self-reported confidence. This finding challenges the common assumption in many studies that self-reported confidence or competency can reliably approximate actual AI-related abilities because, if that assumption held true, one would expect at least a positive correlation between perceived and actual performance. Our result instead supports the presence of an \textit{illusion of competence}, aligning with the findings of \citet{prather2024widening}, who similarly identified a mismatch between learners’ perceived and actual skills when interacting with AI-generated code explanations.

Notably, students’ self-reported confidence shifted from a negative (though non-significant) correlation with their assessment scores before the intervention to a significant positive correlation afterward. This suggests that learners' perceived confidence, or confidence calibration \citep{hattie2013calibration,fisher2021better,muenks2018can}, more realistically reflects their ability after the learning session. These findings indicate that our scaffolded writing practice with AI not only enhances learners' skill acquisition but also supports the development of accurate self-assessment. However, as the focus of this study is on the ability changes rather than perception changes, we only applied limited number of questions to measure a small sample size of students confidence. Future research can more accurately measure whether confidence calibration can be corrected by our training.

\section{Conclusion}
As AI tools become increasingly integrated into everyday learning, their potential to enhance or hinder educational outcomes depends largely on how they are used. While AI holds great promise for supporting learning, improper or uncritical use can undermine it. To unlock AI’s full educational potential and reduce misuse, we first introduced a new concept: pedagogical prompt. This is a type of prompt that guides AI to act as a tutor and generate responses aligned with the student’s current learning stage. We then demonstrated its practical use in the context of early undergraduate computer science education. For the proof-of-concept, we first conducted a formative study with 36 instructors teaching early-stage CS courses in higher education to confirm the classroom needs of teaching pedagogical prompting. Based on instructor insights, we then developed a learning intervention, including an interactive system and a scenario-based learning material set. A user study with 22 CS novices revealed significant improvements in students’ skills in writing pedagogical prompts, as well as increased positive attitudes toward their perception of their prompting ability for learning purposes, and a greater willingness to use pedagogical prompts in their future learning.

Our findings underscore the importance and one effective approach of equipping learners with prompt construction skills to enable more effective self-directed learning. Educators, in turn, play a vital role in fostering meta-cognitive awareness and incorporating pedagogical prompting into instructional design. Meanwhile, AI systems must be designed to scaffold these practices through structured interactions and timely feedback. Looking ahead, we envision broader integration of pedagogical prompting into classrooms through this scenario-based, interactive learning intervention extend beyond computing education. This work offers a foundational step toward re-imagining the learner-AI interaction to promote deeper engagement and greater learner autonomy.

\section{Appendix: Example Implementations of Pedagogical Prompt}

\subsection{Generating Worked Example}

I'm a python beginner having trouble with debugging. The coding problem, my code, and output are as follows:

\textit{[problem description]}

\textit{[current code]}

\textit{[current output]}

Can you act as am intro-level programming tutor and generate a minimal-code example of a different problem that uses a for loop to iterate over indices? Don't give me the solution to the problem.

\subsection{Generating Guiding Questions}
I'm a python beginner having trouble with debugging. The coding problem, my code, and output are as follows:

\textit{[problem description]}

\textit{[current code]}

\textit{[current output]}

Can you act as an intro-level programming tutor and give me 3-4 step-by-step guiding question, with each as a multiple choice question to guide me think through the problem? Don’t ask me the next question until I answered the current question. Don't give me the solution to the problem.

\subsection{Generating In-Context Explanation}

I'm a python beginner having trouble with understanding the problem description. Can you act as am intro-level programming tutor and give me examples of inputs and outputs, and explain how the input becomes the output in the problem context? The problem description is as follow:

\textit{[problem description]}

Don't give me the solution to the problem.

\section{Appendix: Validation Criteria in System Prompts}

\subsection{Problem Type}

\subsubsection{Understand the Task Description}
{\sloppy 
\begin{itemize}
  \item \textbf{\texttt{Mentioning GPT will serve as a tutor of an intro-level programming course}}: \texttt{"The prompt should specify the role of LLM is a tutor for intro programming course. e.g., You are a tutor for intro programming course."}
  \item \textbf{\texttt{Mentioning the problem type that you just chose in the prompt}}: \texttt{"The prompt should mention that [current scenario's character]'s current difficulty is Understand the Task Description. e.g., I need help fixing undesired output."}
\end{itemize}}

\subsubsection{Plan the General Idea of the Solution}
{\sloppy \begin{itemize}
  \item \textbf{\texttt{Mentioning GPT will serve as a tutor of an intro-level programming course}}: \texttt{"The prompt should specify the role of LLM is a tutor for intro programming course. e.g., You are a tutor for intro programming course."}
  \item \textbf{\texttt{Mentioning the problem type that you just chose in the prompt}}: \texttt{"The prompt should mention that [current scenario's character]'s current difficulty is Planning the General Idea of the Solution. e.g., I need help fixing undesired output."}
\end{itemize}}

\subsubsection{Write the Code}
{\sloppy \begin{itemize}
  \item \textbf{\texttt{Mentioning GPT will serve as a tutor of an intro-level programming course}}: \texttt{"The prompt should specify the role of LLM is a tutor for intro programming course. e.g., You are a tutor for intro programming course."}
  \item \textbf{\texttt{Mentioning the problem type that you just chose in the prompt}}: \texttt{"The prompt should mention that [current scenario's character]'s current difficulty is Writing the Code."}
\end{itemize}}

\subsubsection{Fix a Bug with Error Message}
{\sloppy \begin{itemize}
  \item \textbf{\texttt{Mentioning GPT will serve as a tutor of an intro-level programming course}}: \texttt{"The prompt should specify the role of LLM is a tutor for intro programming course. e.g., You are a tutor for intro programming course."}
  \item \textbf{\texttt{Mentioning the problem type that you just chose in the prompt}}: \texttt{"The prompt should mention that [current scenario's character]'s current difficulty is fixing a bug with error message. e.g., I need help fixing undesired output."}
\end{itemize}}

\subsubsection{Fix Undesired Output}
{\sloppy \begin{itemize}
  \item \textbf{\texttt{Mentioning GPT will serve as a tutor of an intro-level programming course}}: \texttt{"The prompt should specify the role of LLM is a tutor for intro programming course. e.g., You are a tutor for intro programming course."}
  \item \textbf{\texttt{Mentioning the problem type that you just chose in the prompt}}: \texttt{"The prompt should mention that [current scenario's character]'s current difficulty is fixing undesired output. e.g., I need help fixing undesired output."}
\end{itemize}}

\subsection{Context Provided}
{\sloppy \begin{itemize}
  \item \textbf{\texttt{Problem Description}}: \texttt{"The prompt must include the description of current coding problem."}
  \item \textbf{\texttt{Current Code}}: \texttt{"The prompt must contain current code snippets. Syntax errors, incompleted code or code too short (e.g. only function definition) should be considered criteria met (because this is how code is presented to them)."}
  \item \textbf{\texttt{Current Output}}: \texttt{"The prompt must show error messages, current results, or actual output from their code. Empty output or error messages should be considered criteria met, if previous code shows significant error from executing (because this is how code is presented to them)."}
\end{itemize}}

\subsection{Learning Method}
\subsubsection{Contextualized Explanation}
{\sloppy \begin{itemize}
  \item \textbf{\texttt{Specify which part you need explanation on}}: \texttt{"Check if the prompt specify which part (e.g., input, output, error message, task description, etc.) need to be explained."}
  
  \item \textbf{\texttt{Use example input and output to walk me through the problem in context}}: \texttt{"Check if the prompt asks for explanation in this problem's context with input and output (e.g., associate the error message to line numbers and code, walk through how to input become the output in this task context)."}
\end{itemize}}

\subsubsection{Example Code on Similar Problem}
{\sloppy \begin{itemize}
  \item \textbf{\texttt{Provide a similar coding snippet that demonstrates the correct use of syntax}}: \texttt{"Check if the prompt mentioned a similar problem as example. "}
  \item \textbf{\texttt{Not the exact same problem as the current one}}: \texttt{"Check if the prompt specifies the example should not be the solution. "}
\end{itemize}}

\subsubsection{Step-by-Step Guiding Questions}
{\sloppy \begin{itemize}
  \item \textbf{\texttt{Specify the chosen strategy: guiding question}}: \texttt{"Check if the prompt explicitly asks to be guiding questions"}
  \item \textbf{\texttt{In a reasonable number of steps (e.g. 3 to 5 steps)}}: \texttt{"Check if the prompt asks for the problem to be broken down into a reasonable number of steps (e.g. 3~5 steps)."}
  \item \textbf{\texttt{Specify each step to be a multiple-choice question to provide clearer guidance and reduce mistakes}}: \texttt{"Check if the prompt asks for a multiple-choice question at each step. This helps guide the process more effectively and reduces the chance of going in the wrong direction."}
  \item \textbf{\texttt{Only ask one question at a time}}: \texttt{"Make sure the prompt doesn't ask the next question before the student answers the current one."}
\end{itemize}}

\subsection{Learner Level}
\subsubsection{Beginner Python Programmer}
{\sloppy \begin{itemize}
  \item \textbf{\texttt{Specifying the guidance is for python beginner}}: \texttt{"Check if the prompt mentions being new to Python or taking first programming course"}
  \item \textbf{\texttt{Asking for simple explanations and syntax}}: \texttt{"Check if the prompt asks for simple or beginner-friendly languages and syntax."}
\end{itemize}}

\subsubsection{Advanced Python Programmer}
{\sloppy \begin{itemize}
  \item \textbf{\texttt{Advanced Knowledge}}: \texttt{"Check if the prompt demonstrates understanding of advanced Python concepts"}
  \item \textbf{\texttt{Technical Language}}: \texttt{"Check if the prompt uses advanced technical terminology correctly"}
  \item \textbf{\texttt{Complex Solutions}}: \texttt{"Check if the prompt shows ability to work with complex problem-solving approaches"}
\end{itemize}}

\subsection{Do Not Provide}

{\sloppy \begin{itemize}
  \item \textbf{\texttt{NOT Include a full solution or complete code}}: \texttt{"The prompt must specifically mention avoiding direct solutions or complete code answers"}
  \item \textbf{\texttt{(Optional) If there are anything else you want to exclude}}: \texttt{"this one is optional, it will always be correct as long as student did not as for the correct answer."}
\end{itemize}}

\section{Declaration of Generative AI and AI-Assisted Technologies in the Writing Process}

\subsection{The use of generative AI and AI-assisted technologies in scientific writing}
Grammarly, Notion AI, and ChatGPT were used in the paper refinement stage to improve the readability and language of the work. All revisions were made on the word or sentence level under human oversight, and we carefully reviewed and edited every part of the content.

\subsection{The use of generative AI and AI-assisted tools in figures, images and artwork}
During the user study of this project (prior to the paper writing stage), the author(s) used OpenAI DALL·E3 to generate visual elements (cartoon characters) for the user study materials. These images were incorporated into comics and were then edited by the author(s) to fit the narrative of the programming scenarios. Visual demonstrations of these materials are included in Figure \ref{fig: example scenarios} and Figure \ref{fig: interface_combined}. This use aligns with elsevier GenAI use policy, as the AI-generated visuals were part of the user study and were not used to manipulate or obscure data, or included in the paper writing process. The tool, version, and process (generating and hand‑editing comics) are described to ensure transparency and reproducibility. After using this tool/service, the author(s) reviewed and edited the content as needed and take(s) full responsibility for the content of the published article.

\bibliographystyle{cas-model2-names}

\bibliography{references}

\end{document}